%% file: main.tex
\documentclass[a4paper,11pt]{article}
\usepackage{jinstpub}

\usepackage[space]{grffile}
\usepackage{dcolumn}
\usepackage{bm}
\usepackage{siunitx}
\usepackage[version=4]{mhchem}
\usepackage{makecell}
\usepackage{tablefootnote}
\DeclareSIUnit\PSI{psi}
\DeclareSIUnit\PSIA{psia}
\DeclareSIUnit\ppb{ppb}
\DeclareSIUnit\ppm{ppm}
\usepackage{xcolor}
\usepackage{pifont}
\hypersetup{
    colorlinks,
    linkcolor={red!50!black},
    citecolor={blue!50!black},
    urlcolor={blue!80!black}
}

\newcommand{\radiopure}{radiopure}
\newcommand{\cthreefeight}{C$_3$F$_8$}

\newcommand{\setpoint}{setpoint}
\newcommand{\rntwotwotwo}{$^{222}$Rn}
\newcommand{\potwoten}{$^{210}$Po}
\newcommand{\pbtwoten}{$^{210}$Pb}
\newcommand{\pbtwofourteen}{$^{214}$Pb}
\newcommand{\rightsideup}{Right-Side-Up}

\newcommand{\ignore}[1]{}

\title{The PICO-40L detector} 

\input{authors}

\date{\today}

\abstract{
The PICO collaboration has designed and constructed a novel bubble chamber to search for evidence of Weakly Interacting Massive Particles, a leading dark matter candidate. The PICO-40L detector is the successor to PICO-60, and uses a fully redesigned system for pressurization of the active volume after a bubble nucleation event. Pressure in PICO-40L is transferred to the target fluid through a set of two nested synthetic quartz vessels connected by bellows \emph{below} the active volume. This is in contrast to the design of earlier detectors in which the pressure was transferred through a buffer fluid atop the target fluid, within a single synthetic quartz vessel connected to a bellows \emph{above} the active volume. Removing the buffer fluid removes the potential for backgrounds arising from the interface between the two fluids, and moving the bellows below the target fluid suppresses potential backgrounds from particulates originating from the bellows. The newly designed detector concept of PICO-40L is referred to as the \rightsideup{} (RSU) design and requires a dedicated thermal system to keep the lower region of the detector cold while the upper region is stably superheated. The main goal of PICO-40L is to establish the performance characteristics of the RSU detector concept, which will inform the operating configuration of the final generation PICO dark matter detector, PICO-500, which is currently under construction at SNOLAB. This paper reviews the design, installation, and performance of the PICO-40L detector.}

\begin{document}
\maketitle
\flushbottom


\section{Bubble chambers for dark matter detection}

Astrophysical measurements over the last few decades have refined our knowledge of the existence and nature of dark matter. The strongest evidence for dark matter comes from a wide variety of cosmological and astrophysical observations, such as precise measurements of the Cosmological Microwave Background~\cite{Planck}. Although the nature of dark matter remains unknown, many theories beyond the Standard Model of particle physics predict massive and neutral particles, with the most favoured dark matter particle candidates being Weakly Interacting Massive Particles (WIMPs)~\cite{BERTONE2005279}. Direct detection experiments aim to detect WIMPs via scattering off of target nuclei within a detector on Earth, resulting in low-energy (keV-scale) nuclear recoil signals. Expected rates for WIMP-scattering events are orders of magnitude lower than natural radioactivity, requiring well-shielded detectors made of \radiopure{} materials operated deep underground with large exposures.

The PICO experiment aims to directly detect WIMPs using a liquid target of \cthreefeight{}. $^{19}$F has a natural isotopic abundance of 100\% and a high spin enhancement factor from its single unpaired proton~\cite{Tovey:2000mm,Bednyakov}, making fluorine-rich targets such as \cthreefeight{} ideal for dark matter searches in the spin-dependent sector. The low mass number of fluorine also leads to high sensitivity in the low-mass WIMP range of approximately \SIrange{1}{50}{GeV/$c^2$}, an area of interest in dark matter experiments~\cite{DMReviewBozorgnia}. The operating principle of a bubble chamber configured for dark matter detection relies on putting the target fluid in a superheated state, which enables a small, localized energy deposition within the target fluid to form a macroscopic bubble. Acoustic waves caused by the nucleation event and bubble growth, as well as images of the growing bubble, are captured and read out, after which the pressure in the detector is rapidly increased to collapse the expanding bubble and return the target to a fully liquid state. In the spin-dependent (proton) sector, the superheated bubble chamber has been among the detector technologies at the forefront, with the most stringent limits set by the PICO and LZ collaborations~\cite{PICO:2015pux,PICO:2019vsc,LZ:2025}. In this context, PICO detectors benefit from several key features intrinsic to bubble chambers. Because the energy deposited by electronic recoils (from gamma radiation) are generally insufficiently localized to nucleate a bubble, PICO detectors are less sensitive to this class of background by orders of magnitude compared to most other types of detectors, which allows for low, background-free detection thresholds (order of keV)~\cite{PICO:2019rsv}. Additionally, the power of acoustic discrimination between nuclear recoils and the main source of radioactive background --- alpha particles --- enables PICO detectors to powerfully reject internal sources of radioactivity~\cite{COUPP:2011}. Another feature is the ability of PICO chambers to observe neutrons which scatter multiple times in the detector, which are easily distinguished from single nuclear recoils and removed. Finally, PICO detectors can be operated with a variety of targets fluids: targets containing elements with low atomic mass number, such as the hydrogen-abundant C$_2$H$_2$F$_4$, lead to optimal momentum transfers for low-mass WIMPs in the GeV/$c^2$ mass range and significantly increases the sensitivity to sub-GeV/$c^2$ dark matter, while targets containing elements with high atomic mass number, such as the iodine-containing CF$_3$I, offer good sensitivity to spin-independent interactions. 

\section{Description of PICO-40L}
\subsection{Location and design concept}
\label{sec:loc_dc}

The PICO-40L experiment is located at the SNOLAB underground laboratory in the Vale Creighton mine in Greater Sudbury, Ontario, Canada. The surface profile of the mine site is flat, and the \SI{2072}{m} overburden gives approximately \SI{6000}{m} water-equivalent shielding, among the highest of any laboratory on Earth~\cite{Woodley:2024eln}, resulting in a cosmic-ray muon flux of \SI{0.27}{m$^{-2}\mathbin{\cdot}$day$^{-1}$}~\cite{SNO:2009oor}. The PICO-40L experiment is located in the Ladder Lab area, which was built to accommodate medium-sized experiments~\cite{SNOLAB_Handbook}.

\begin{figure}[htbp]
	\begin{center}
	\includegraphics[totalheight=0.5\textheight]{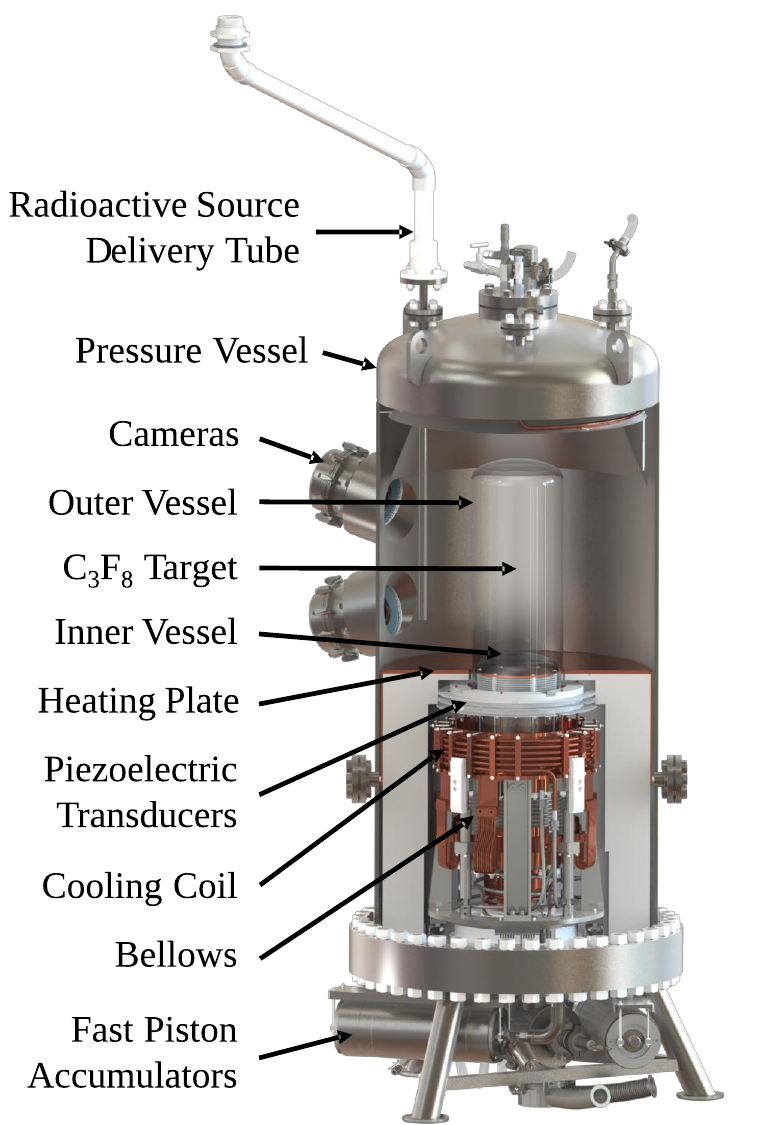}
    \includegraphics[totalheight=0.45\textheight]{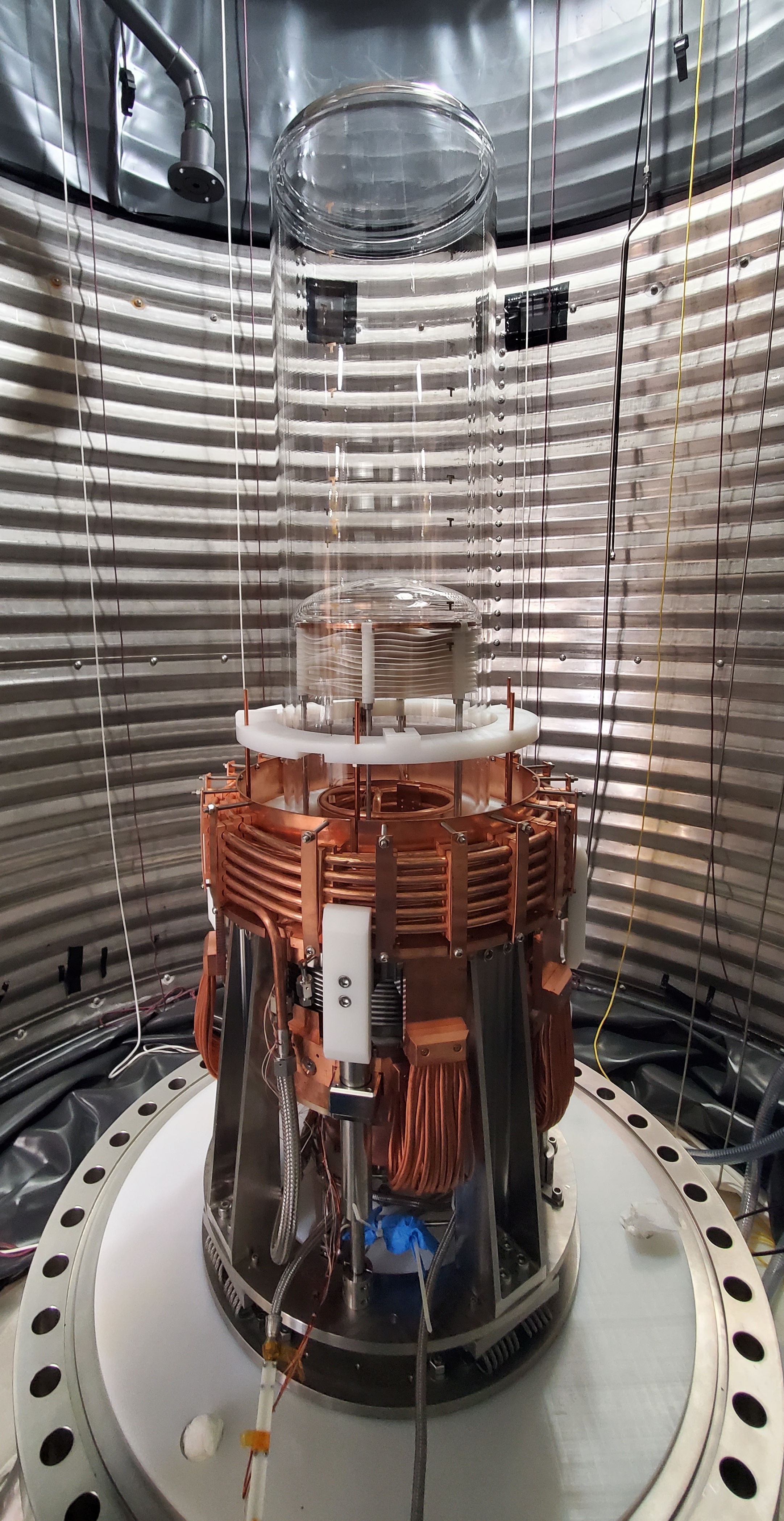}
	\end{center}
	\caption{(Left) Schematic side view of the PICO-40L pressure vessel (PV) with the major detector components labelled. The synthetic quartz vessels house the \cthreefeight{} target. Two of the four camera ports are visible on the left of the PV. The heating plate and cooling coil maintain the temperatures of the two thermal zones. (Right) The PICO-40L inner assembly, bellows region cooling coil, and other PV-internal components, partially assembled inside the water tank.}
	\label{fig:Detector}	
\end{figure}

Figure~\ref{fig:Detector} shows a schematic of the PICO-40L detector (left) along with a photo of several internal components of the detector after assembly (right). The core of the detector is the two nested synthetic quartz vessels (the smaller inner vessel and the larger outer vessel), between which the \cthreefeight{} target fluid is contained. The superheated portion of the target fluid is defined as the active volume. The open ends of the vessels are attached to opposite sides of the stainless-steel bellows below the active volume. The two quartz vessels along with the bellows are collectively known as the "inner vessel assembly". The cooling coil in the bellows region is designed to prevent the target fluid in this region from becoming superheated, thereby suppressing bubble nucleation, while the heating plate around the top of the inner vessel is designed to keep the target fluid above it stably superheated. The inner vessel assembly, cooling coil, and heating plate are contained within a stainless-steel pressure vessel (PV) rated for a maximum operating pressure of \SI{1517}{kPa} (\SI{220}{\PSIA}). The PV is filled with Crystal Plus 70FG~\cite{mineral_oil} mineral oil, chosen for its optical clarity and stability at cold temperatures, which is used as a hydraulic fluid. The pressure of the mineral oil is controlled by an external hydraulic control system, and is transmitted to the target fluid through the movement of the bellows. Additionally, the mineral oil acts as a thermal bath, allowing for stable control of the temperature of the active volume. The active volume is viewed through four cameras, each of which is encircled by a custom LED ring to provide illumination. Each camera and LED pair is housed in an individual camera port on the side of the PV. To improve the overall illumination of the active volume, a pair of custom stainless-steel cones (called the ``retroreflector cones'') are covered in retroreflecting foil and mounted on the top of the inside of the PV. The back of the PV is additionally covered in retroreflecting foil. Acoustic signals from bubble nucleation events are read out by piezoelectric transducers near the base of the outer vessel. The PV is housed within a stainless-steel tank filled with \SI{20}{tonnes} of ultrapure water (UPW) for thermal stability and shielding from external radiation.

A significant design difference between PICO-60 and PICO-40L is that the PICO-40L inner vessel assembly is mounted in the \rightsideup{} (RSU) geometry, allowing for a buffer-free inner volume with two temperature zones; an active superheated zone and an inactive cold zone. The RSU design concept has been previously employed on smaller bubble chambers filled with xenon~\cite{Baxter:2017ozv} and fluorocarbons~\cite{Bressler:2019xhk}, and will be used in the future larger PICO-500 experiment~\cite{Malin:thesis} and the Scintillating Bubble Chamber (SBC) experiment~\cite{SBC:universe9080346}.

\begin{figure}[htbp]
    \centering
    \includegraphics[width=\textwidth]{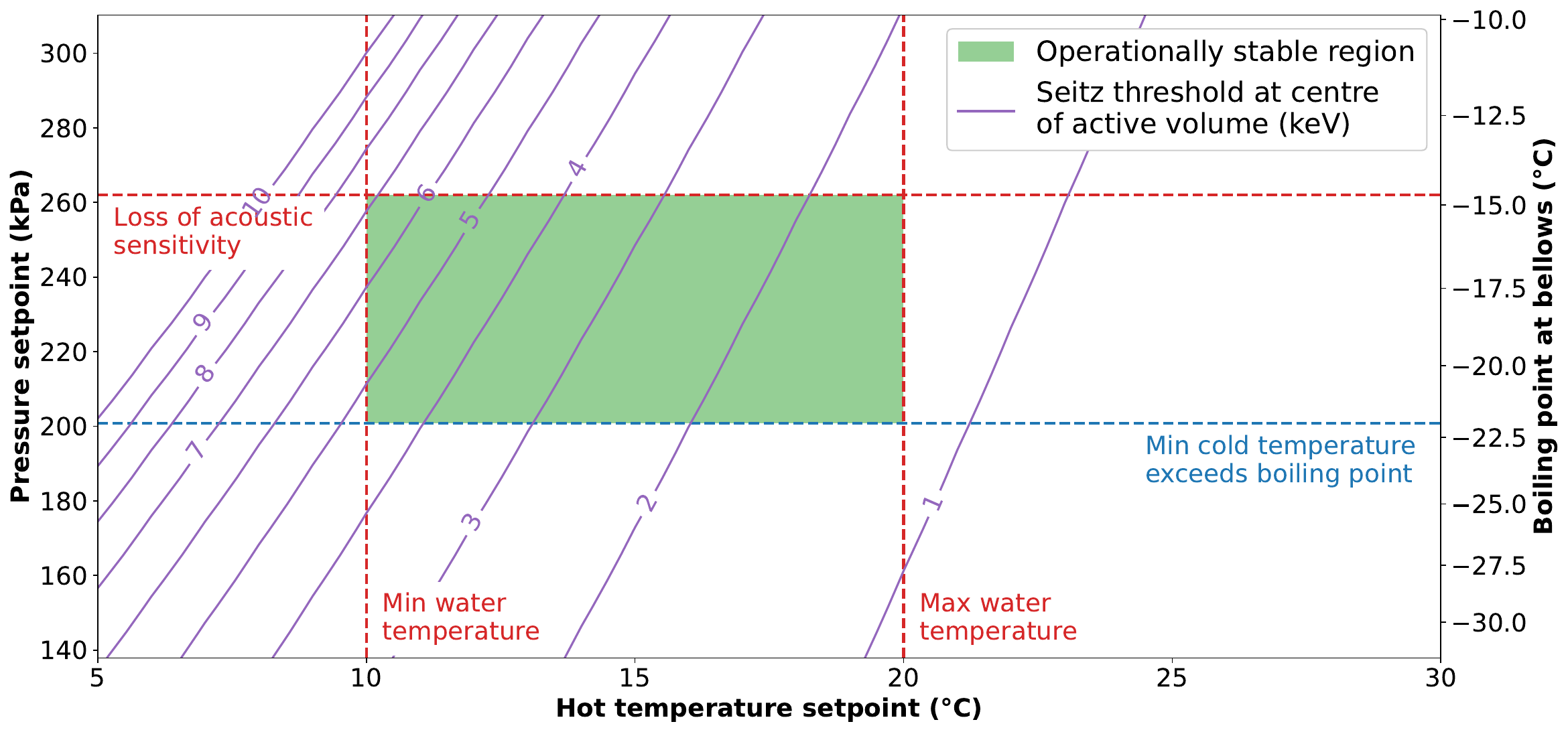}
    \caption{The space of available pressure and temperature \setpoint{}s for the operation of PICO-40L. The x-axis is the temperature \setpoint{} of the upper (hot) thermal zone, and the left y-axis is the expanded pressure \setpoint{}. The right y-axis gives the \cthreefeight{} boiling point in the bellows region at the corresponding expanded pressure \setpoint{}. The purple solid lines are iso-threshold lines in units of keV at the centre of the active volume. The green shaded region depicts the range of \setpoint{}s over which the detector is designed to operate stably.}
    \label{fig:parameter_space_kpa}
\end{figure}

The PV volume is divided into two separate thermal zones. The upper zone is regulated at an operating temperature appropriate to establish dark matter sensitivity in the superheated fluid when the chamber is expanded to operating pressure. This temperature is typically of order \SI{13}{\degreeCelsius}. The temperature \setpoint{} of this upper zone can be adjusted, along with the operating expanded pressure, to precisely control the Seitz~\cite{SEITZ} threshold for bubble nucleation in the chamber. The range of temperatures and pressures around which the detector has been designed to operate is shown graphically in Figure~\ref{fig:parameter_space_kpa}. Since the Seitz threshold depends on both the temperature and the pressure, the range of operating temperature and pressure \setpoint{}s that are available in the upper region allow for the selection of a fixed Seitz threshold along a curve in temperature-pressure space. It has been shown that the probability of electronic recoil events to induce a bubble nucleation deviates from the predictions of the Seitz model, allowing the operating \setpoint{}s of the detector to be tuned to minimize sensitivity to this class of background event without a corresponding decrease in sensitivity to nuclear recoils~\cite{PICO:2019rsv}. The green shaded region indicates the range of operational \setpoint{}s where the detector is designed to operate stably. The upper boundary of this region is set by the maximum pressure at which acoustic separation between alpha events and nuclear recoil events is expected to be possible (see Section~\ref{sec:acoustic_discrimination}), and the lower boundary is set by the minimum pressure at which the boiling point in the bellows region remains above the minimum temperature achievable by the detector cooling system in that region. The region is bounded on the left and right by the minimum and maximum temperatures achievable by the water tank circulation system (see Section~\ref{sec:thermal_systems}). The solid purple lines are iso-threshold curves, i.e. curves in pressure and temperature \setpoint{} space with equivalent Seitz threshold. Since the threshold will vary from the top to the bottom of the active volume due to the varying hydrostatic pressure, the iso-threshold curves are given at the centre of the active volume, defined nominally as the point \SI{225}{\mm} below the top of the cylindrical region of the outer vessel and equidistant from the cylindrical region walls.

The lower zone is maintained at a temperature of nominally \SI{-22}{\degreeCelsius}, and the expansion pressure is selected such that the \cthreefeight{} within it is in the non-active (not superheated) state over the entire operational pressure range. This ensures that bubble nucleation events rarely occur outside the view of the cameras, particularly at the quartz vessel flanges, seals, stainless-steel components, and plumbing connections, where nucleation sites would otherwise exist. The inner vessel assembly, external hydraulic controls, thermal control system, photography, and instrumentation are described in detail in the following sections.

The PICO-40L design is an evolution of two previous PICO fluorocarbon bubble chambers, PICO-2L~\cite{PICO:2015yox} and PICO-60~\cite{PICO:2015pux,PICO:2017tgi}. The active volume of the PICO-40L chamber is \SI{41.7\pm0.1}{L} of \cthreefeight. Although similar in size to PICO-60 (\textasciitilde\SI{40}{L} for PICO-40L compared to \textasciitilde\SI{37}{L} for PICO-60), the design of PICO-40L improves the dark matter sensitivity by a factor of five. The improved sensitivity is primarily achieved by reducing neutron backgrounds through materials selection and additional shielding provided by the mineral oil in the larger PV.

\subsection{Inner vessel assembly}
\label{sec:inner}

\subsubsection{Vessel design}
The role of the inner vessel assembly is to provide containment for the target fluid within the hydraulic fluid. The inner assembly must be optically clear to allow for observation of bubble formation, sufficiently smooth to limit spurious nucleation from surface roughness, allow for liquid volume changes with minimal pressure differential so that the pressure of the hydraulic environment can be coupled to the active volume, provide for fluid plumbing connections that are not sources of spurious bubble nucleation, and be sufficiently free of alpha-emitting radionuclides. Two nested synthetic quartz vessels (HSQ 900 from Heraeus) and a flexible stainless-steel bellows provide containment and a variable volume for the bubble chamber as shown in Figure~\ref{fig:Detector}. Two fill ports on the bellows are connected to stainless-steel lines which run out of the PV through a dedicated feedthrough flange, and remain in place during operation, which allows the active mass of the target fluid to be increased or decreased, or the target fluid to be changed, with minimal disassembly of other components. The outer vessel has an outside diameter of \SI{300}{mm} and overall length of \SI{1001}{mm}; the inner vessel has outside diameter of \SI{272}{mm} and overall length of \SI{673}{mm}. Both vessels have a wall thickness of \SI{5}{mm}. The total volume of \cthreefeight{} contained is \SIrange{44.9}{51.5}{L}, depending on the extension of the bellows, of which \SIrange{38}{45}{L} is in the active volume. A small \textasciitilde\SI{1}{cm} annulus between the inner and outer vessels allows for fluid flow to the external plumbing while preventing any contact between the vessels and minimizing thermal convection between the cold region of the bellows and the active volume. The inner vessel assembly is mounted on a set of seismic isolation springs to suppress acoustic signals cause by motion in the lab surrounding the detector. The position of the inner vessel assembly relative to the PV is monitored by a Gefran RK-4-150-N magnetostrictive position sensor.

\subsubsection{Bellows design}
The bellows assembly is made from a Hyspan 7577S hydroformed stainless-steel bellows welded to stainless-steel flanges. The bellows flanges provide machined surfaces to ensure a tight seal against the outside of the clamping flanges, and are equipped with mounting points that serve as connections to the mechanical guide rod assembly, providing support for the motion. The bellows can be seen below the quartz vessels in Figure~\ref{fig:Detector} (right). The upper bellows flange mounts directly to the support frame, while the lower flange mounts to three linear bearings that ride on a hardened shaft to ensure that the bellows flanges remain parallel. The bellows has an inside diameter of \SI{30}{\cm} to match the outer quartz vessel flange, an effective area of \SI{800}{\cm$^2$}, and length of \SI{24.5}{\cm}, of which \SI{12.7}{\cm} is corrugated, allowing for a maximum stroke of \SI{\pm5}{\cm}. The range of motion is restricted to \SI{\pm4.1}{\cm} by mechanical stops on the guide tubes, sufficient to accommodate thermal expansion of the \cthreefeight{}; from the operating temperature (around \SI{13}{\degreeCelsius}) to \SI{25}{\degreeCelsius}, the \cthreefeight{} volume increases by \SI{3.0}{\L}, corresponding to a bellows movement of \SI{4}{\cm}. Volume changes during operation are smaller, with the bellows moving by \textasciitilde\SI{0.5}{\cm}. The bellows position is monitored by a Gefran RK-4-150-N magnetostrictive position sensor.

\subsubsection{Vessel seal}
\label{sec:vessel_seal}
\begin{figure}[htbp]
    \centering
    \includegraphics[width=0.95\textwidth]{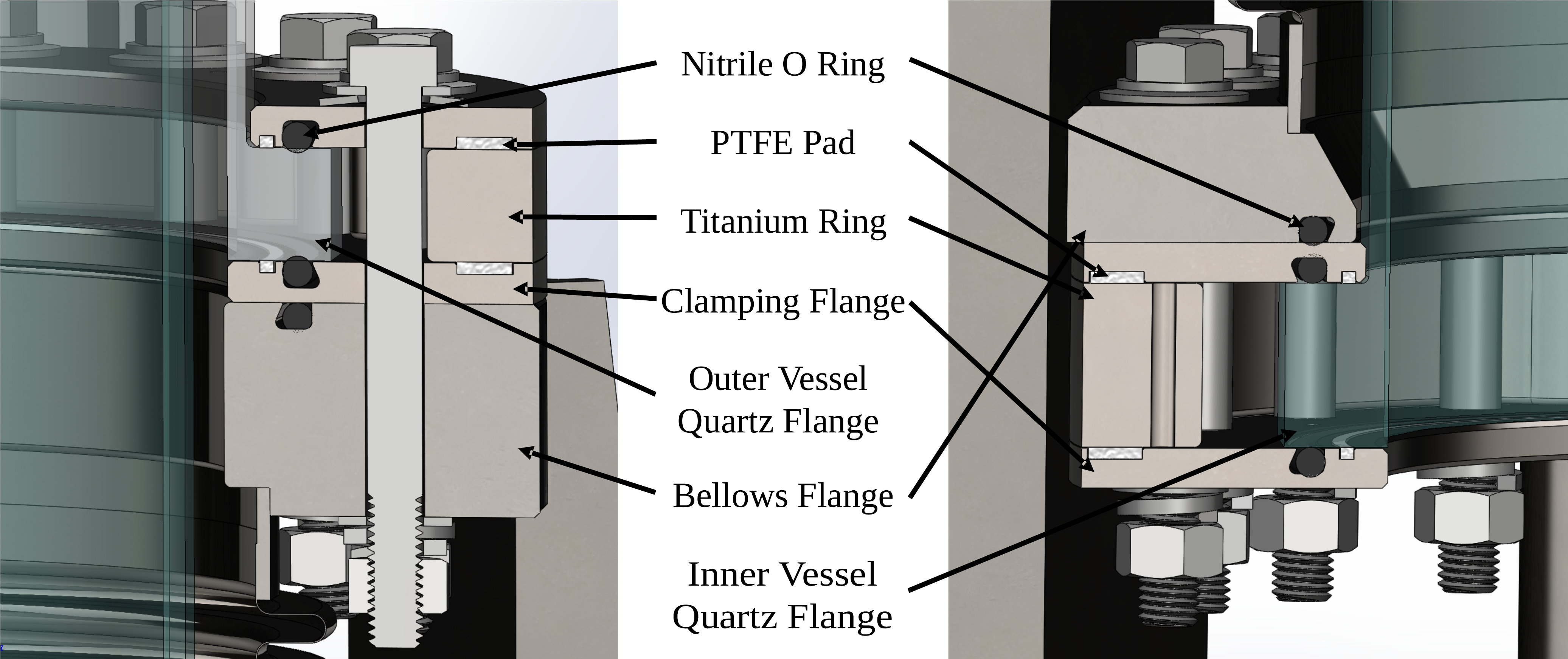}
    \caption{Outer (left) and inner (right) vessel seals. The nitrile O-rings provide the seal between the inner and outer volumes, while the PTFE pads prevent the quartz vessel from contacting the stainless-steel clamping flanges over the full temperature range anticipated during construction and operation. The titanium ring around the quartz flange is used to more closely match the thermal expansion and contraction of the quartz material.}
    \label{fig:seal}
\end{figure}

The selection of a sealing material for bubble chamber use is constrained by the following requirements: the seal needs to be compatible with the hydraulic fluid outside the inner vessel assembly (mineral oil) and the target fluid inside (\cthreefeight{}), to work throughout the full temperature range of operation (\SIrange{-25}{30}{\degreeCelsius}), to allow a pressure differential of about \SI{138}{kPa} between the inside and the outside, and to have low radon emanation and permeation properties. For PICO-40L, a nitrile O-ring was selected, as it is the only elastomer identified that allows operation below \SI{-25}{\degreeCelsius} and is compatible with \cthreefeight{} and mineral oil. The choice of nitrile as a sealing material has a significant impact on the background rate of PICO-40L, having a radon emanation rate on the order of \SI{3e-7}{Bq$\mathbin{\cdot}$day$^{-1}\mathbin{\cdot}$cm$^{-1}$}. Alone, the rather soft elastomer could not prevent contact between the steel and quartz surfaces in upset conditions, therefore all seals were framed by polytetrafluoroethylene (PTFE) pads, designed so that the elastomer seal is properly compressed at all times, while contact between the steel and quartz surfaces is prevented. This is shown in Figure~\ref{fig:seal}, both for the seal of the outer (left) and the inner (right) quartz vessels. To allow the seal to function during temperature changes, such as the initial cool-down, a titanium ring around the quartz flanges is used to more closely match the thermal expansion of the quartz. 

\subsection{Pressure control and hydraulic system}

\begin{figure}[htbp]
	\begin{center}
		\includegraphics[width=0.9\textwidth]{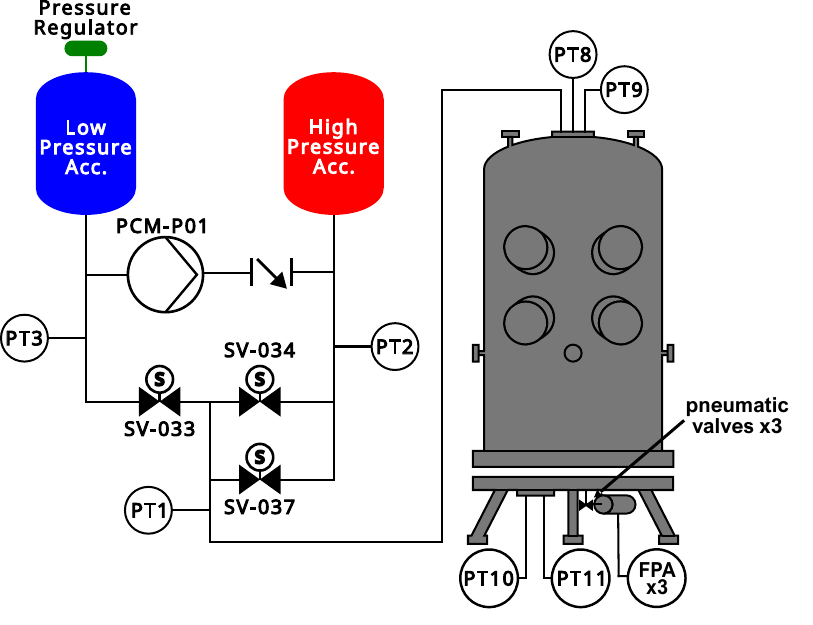}
	\end{center}
	\caption{Simplified pressure system schematic. To expand the detector at the beginning of a cycle, the three pneumatic valves and solenoid valves SV-034 and SV-037 are closed, and SV-033 is opened, allowing oil to the low pressure side of the system and lowering the pressure in the inner vessel assembly. To compress the detector, the valve states are reversed, preventing further boiling of the target fluid. After compression, oil is moved from the low pressure side to the high pressure side using the PCM-P01 hydraulic pump. The numbered pressure transducers (PTs) monitor the pressure at various points in the system. Figure from~\cite{Moore:thesis}.}   
	\label{fig:Pressure}	
\end{figure}

The design of the pressure control system for PICO-40L, a simplified view of which is shown in Figure~\ref{fig:Pressure}, is similar to previous PICO detectors~\cite{PICO:2015pux,PICO:2015yox}. The addition of three custom ``fast'' piston accumulators (FPAs) mounted on the base of the PV allows for rapid compression within $\mathcal{O}$(\SI{100}{\ms}) of receiving a compression signal, faster than would be possible for such a large chamber with hydraulic pumping alone. Fast compression of bubbles reduces the time required to cycle the chamber back to a stable superheated state. The three FPAs are custom \SI{9}{\L} accumulators, the bodies for which are manufactured by KOCSIS Technologies. A normally closed pneumatic valve (ASCO model 8290B048) is welded to the output of each accumulator, isolating it from the PV during expansions.

The pressure control system consists of three major components: the stainless-steel PV which encapsulates the inner vessel assembly, the three FPAs mounted to the base of the PV, and the hydraulic system, which contains a hydraulic pump, a programmable logic controller (PLC, National Instruments cRIO-9074), and all other necessary components to control the pressure in the PV.

\begin{figure}[htbp]
    \centering
    \includegraphics[width=\textwidth]{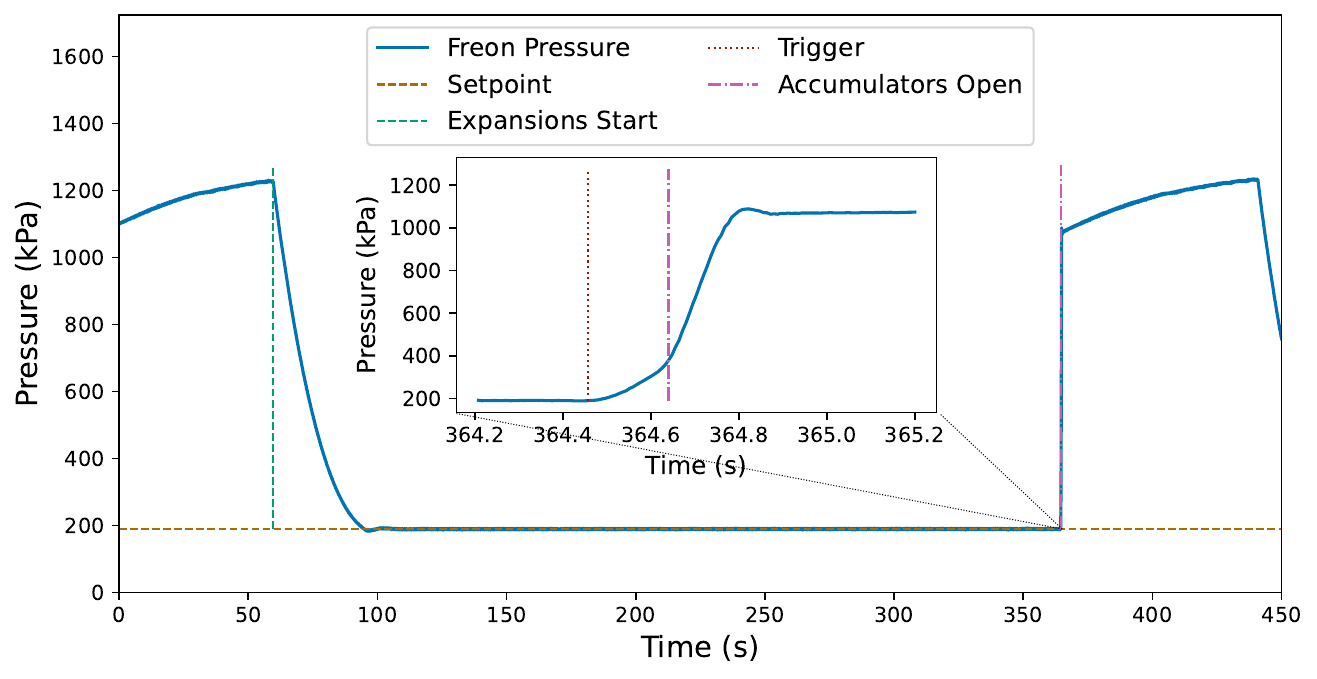}
    \caption{An example expansion cycle for an event in PICO-40L, with closeup of the trigger region inset. The detector is in the superheated configuration when the pressure is at the low \setpoint{}, from t$\approx$\SI{100}{\s} to t$\approx$\SI{360}{\s}. The FPAs open rapidly after a trigger is received, recompressing the chamber and allowing the cycle to begin again.}
    \label{fig:pressure_cycle}
\end{figure}

A typical expansion cycle is illustrated in Figure~\ref{fig:pressure_cycle}. When inactive, the detector is held in the compressed state, with the target fluid held around \SI{1220}{kPa}, well above its vapour pressure at the operating temperature. For the planned operating temperature range of \SIrange{12}{18}{\degreeCelsius}, the vapour pressure of \cthreefeight{} is between \SIrange{620}{758}{kPa}. In the compressed state, solenoid valves SV-034 and SV-037 (Omega Engineering model SV128 and SV131, respectively) are open, allowing mineral oil to flow between a \SI{19}{\L} KOCSIS Technologies accumulator (model A60-1155-3K-B-20S), called the ``high pressure'' accumulator (HPA), and the PV. In this state, the three pneumatic valves controlling the flow between each of the three FPAs and the PV are also open.

To transition to the expanded state, in which the target fluid is below its vapour pressure and therefore superheated, the HPA and FPAs are isolated from the PV by closing the three pneumatic valves (one for each FPA) as well as SV-034 and SV-037. Solenoid valve SV-033 (Omega Engineering model SV128) is opened, allowing mineral oil to move from the PV to a \SI{9}{\L} Parker bladder accumulator (model BA02B3T01P2), called the ``low pressure'' accumulator (LPA), causing the pressure to drop to the desired \setpoint{}. The pressure in the air side of this accumulator is finely controlled via a Proportion-Air air regulator (model DQPV1TFEE100AXL) connected to the compressed air system at SNOLAB. In this configuration, the detector is sensitive to particle interactions. Once the PLC receives a signal to recompress, the detector is returned to a compressed state by closing the valve to the LPA (SV-033) and opening the valves to the FPAs (three pneumatic valves) and HPA (SV-034 and SV-037). As the net effect of an expansion cycle is to transfer oil from the HPA to the LPA, oil is returned to the HPA while in the compressed state with PCM-P01, a Liquiflo H-series gear pump (model H7NS6PEEU020001), bringing the pressure back up to \SI{1220}{kPa}. After every compression, the chamber is held above the \cthreefeight{} vapour pressure for at least \SI{60}{s}, and after every tenth compression, the chamber is held above the \cthreefeight{} vapour pressure for at least \SI{90}{s}. This is done to ensure that any bubbles in the detector are fully collapsed and the \cthreefeight{} is stable before the subsequent expansion.

The pressures in all regions of the hydraulic system are monitored through a combination of eight pressure transducers (PTs). The three transducers on the cart (PT1, PT2, and PT3), all Omega Engineering model PX319-300A-5V, monitor the pressure at the output of the cart, the high pressure accumulator, and the low pressure accumulator. One Noshok model 100-300-1-1-2-7 and one Setra model 225G250PAD41106 pressure transducer are located in each of two positions; at the top (PT8 and PT9) and bottom (PT10 and PT11) of the PV. The two different models were used at each location for redundancy and to protect against unexpected failure. A final transducer, a Setra model 225G250PAD41106, is in direct contact with the target fluid in the cold region (not shown in Figure~\ref{fig:Pressure}), and serves as the primary measurement of active volume pressure for calculations of the Seitz threshold.

\subsection{Thermal systems}
\label{sec:thermal_systems}
\begin{figure}[htbp]
    \centering
    \includegraphics[width=0.9\textwidth,
    pagebox=mediabox]
    {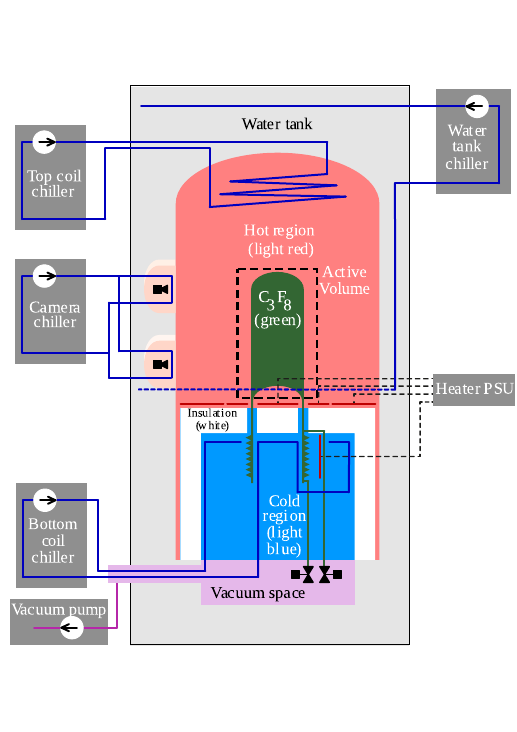}
    \caption{Schematic of the thermal system for PICO-40L. The detector is divided into two thermal zones, separated by an HDPE insulating layer. The active volume is the \cthreefeight{} (green) surrounded by the superheated hot zone (light red), while bubble nucleation is suppressed in the \cthreefeight{} surrounded by the non-superheated cold zone (light blue). A system of cooling loops and heating plates (dark blue and dark red respectively) maintain the temperatures of the two zones.}
    \label{fig:Thermal_schematic}
\end{figure}

The thermal system is shown in Figure~\ref{fig:Thermal_schematic}, and has been redesigned from that of previous chambers to accommodate the RSU detector geometry and to improve temperature control. The cold region (light blue) is insulated from the hot region (light red) by high-density polyethylene (HDPE, white). The insulation was fabricated in the radon-reduced machine shop at the University of Alberta~\cite{Aksel:LRCtalk,Aksel:poster} to minimize the radioactivity of components close to the active volume. The cold region ($\approx$\SI{-22}{\degreeCelsius}) maintains a non-active state of the \cthreefeight{} in contact with stainless-steel components, while the hot region ($\approx$\SI{13}{\degreeCelsius}) keeps the target fluid superheated.

The cold region is maintained by the bottom coil chiller (SP Scientific RC-210) circulating Syltherm XLT chiller fluid through the cooling coil in the bellows region of the detector. The cooling coil has two components: the inner coil is located within the downward-facing inner vessel, while the outer coil fits around the outside of the bellows and is connected to the HDPE. The cooling lines are fed into the water tank through conduits that are kept under vacuum. The output of the chiller first goes through the cold housing mounted under the detector. Inside the housing, the pipe is thermally coupled to a copper plate before feeding the inner cooling coil. This plate keeps the \cthreefeight{} in the surrounding pipes and valves cooled to suppress boiling in this region. The output of the inner coil feeds directly into the outer coil, before returning to the chiller.

The camera chiller (Polyscience 5360T11A1110C) feeds into the four enclosures and is in thermal contact with the cameras and LED rings. This helps to remove excess heat from these heat-generating components and suppresses rotational asymmetry in the thermal profile. The top coil chiller (NESLAB RTE 740) is used to prevent rising heat from creating a temperature gradient in the mineral oil. This chiller feeds the top copper coil that is mounted to the upper retroreflector cone. 

The PICO-40L water tank is double-walled stainless steel, \SI{3.72}{m} high and \SI{2.88}{m} in diameter. The space between the two tank walls is filled with \SI{10}{\cm} of insulation which helps to maintain the interior of the water tank at a stable temperature. The water tank chiller (Lytron RC0995) maintains temperatures around the PV lower than the ambient lab temperature, allowing the mineral oil and \cthreefeight{} to remain at the operating temperature \setpoint{}. The water tank circulation system distributes UPW chilled to \SI{12}{\degreeCelsius} through a sparge ring encircling the PV, a few centimetres above the top of the PV, approximately halfway between the PV outer wall and the water tank inner wall. The chilled water returns to the circulation loop through a second sparge ring with a similar diameter situated at approximately the same height as the heating plates inside the PV. This configuration keeps the top region of the water tank at a temperature close to the hot region \setpoint{}, while allowing the bottom region of the water tank to remain colder, facilitating thermal control inside the PV. The chilling loop also contains an ultraviolet sterilizer (Sterilite V1Q-S5QPA) to suppress biological growth, a polypropylene filter (Pentek P1 with a 1-micron cartridge), and an ion exchange bottle (Regeneration Services UCUY3700). A water sensor (WaterBug WB-350) is deployed on the lab floor outside the water tank, and the water circulation system is configured to automatically disable if the sensor detects water. The temperature, flow rate, resistivity, and leak status of the system are monitored by the detector control system, described in Section~\ref{sec:DAQ}.

Inside the PV, the bottom of the hot region is controlled by heating the copper plates using heater wire, which is embedded into the top layer of HDPE and held in place by Kapton tape. The outer copper plate heats the mineral oil volume surrounding the quartz vessels, while the inner copper plate keeps the mineral oil inside the top of the inner vessel warm. To prevent convection of mineral oil between the cold and hot regions in the small space between the outer vessel and the cold region insulation, the piezoelectric-transducer-housing HDPE ring (described in Section~\ref{sec:acoustic_sensors}) includes HDPE fins that face away from the vessel surface. These fins press into the top layer of the HDPE blocks, preventing flow between regions while allowing flexibility. The final heater shown in Figure~\ref{fig:Thermal_schematic} connects to the \cthreefeight{} fill line and is only used during the filling of the inner vessel assembly as described in see Section~\ref{sec:fluid_fill}. Throughout the detector there are 54 resistance temperature detectors (RTDs), which add redundancy in monitoring the temperature in each region.

The PICO-40L thermal system was simulated using a 2-dimensional axisymmetric cross-section of the detector in COMSOL~\cite{COMSOL}, out to the edge of the water tank, based on a simplified version of the detector geometry. The simulation allows convection in the hydraulic fluid and models the fluid properties of the \cthreefeight{}, hydraulic fluid, and water, along with the thermal properties of all materials in the detector. Boundary conditions are fixed by the heating and cooling elements in the system. The model omits a layer of polystyrene insulation around the base of the PV, which was a late modification to the detector.

\begin{figure}[htbp]
    \centering
    \includegraphics[totalheight=0.5\textheight]{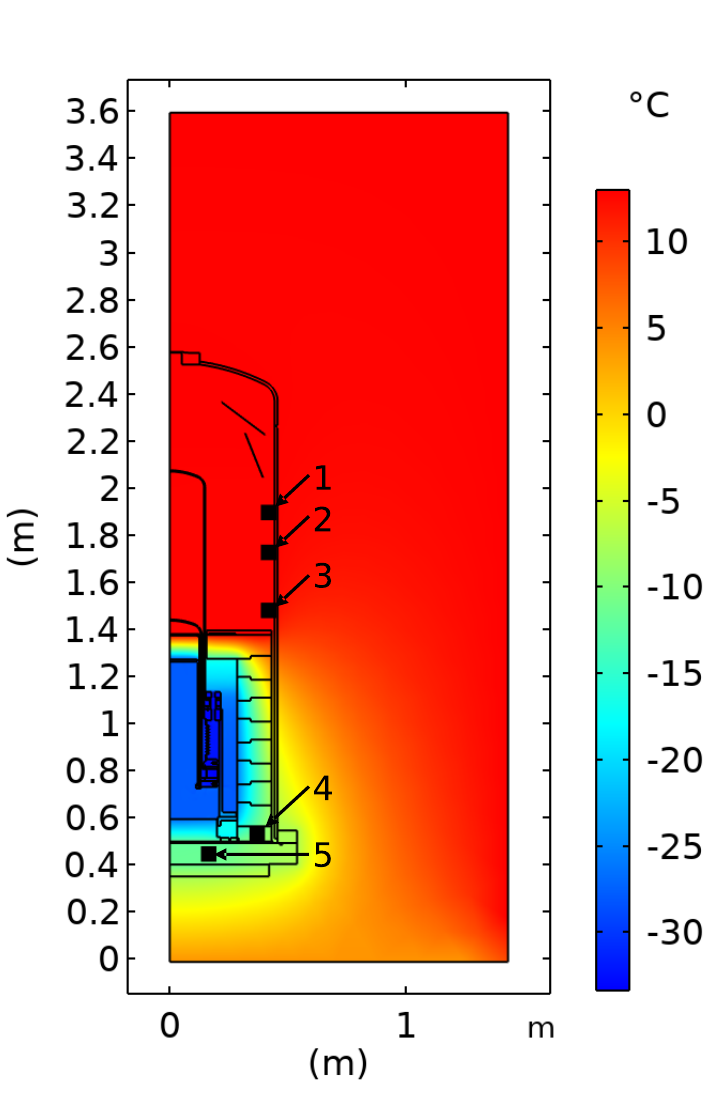}
    \caption{Temperature profile of a 2-dimensional axisymmetric simulation using a simplified model of the PICO-40L detector, after reaching equilibrium. The numbered black squares correspond to locations of RTDs in the detector, and a comparison to commissioning data is shown in Table~\ref{tab:temperature_comparisons}. All simulated temperatures show a discrepancy $\leq$~\SI{4}{\degreeCelsius} compared to the measured temperatures.}
    \label{fig:Thermal_simulation}
\end{figure}

The thermal profile of the simulation solution after reaching equilibrium is shown in Figure~\ref{fig:Thermal_simulation}. The locations of the numbered black squares in the model correspond to RTDs in the detector, and the temperatures in each were compared during a week-long uninterrupted commissioning run, from 21 December 2023 to 28 December 2023. The locations were selected to be far away from thermal gradients and boundary conditions. The minimum, mean, and maximum temperatures of the active volume in the simulation were recorded, as these can be used to constrain the threshold range of the detector, since direct instrumentation of the active volume is not possible in the physical detector. All temperatures in the model were changing at a rate of less than one degree per day, and were well fit by exponential decay functions, allowing the steady state temperatures to be calculated. Table~\ref{tab:temperature_comparisons} shows the extrapolated simulation temperatures and measured temperatures in the detector. The simulated temperatures show good agreement with measured temperatures in the hot region and show a discrepancy $\leq$~\SI{4}{\degreeCelsius} compared to the measured temperatures in the cold region. The uncertainties in the measurement values are propagated from the RTD calibration uncertainty. The discrepancy in temperature between calibration measurement and the simulation in the cold region is driven primarily by the absence of a polystyrene insulating layer in the model. The simulation shows that the heating and cooling system is capable of maintaining the separate thermal zones at the temperatures required for operation.

\begin{table}[htbp]
    \centering
    \caption{Comparison of measured and simulated temperatures in various regions of PICO-40L.}
    \vspace{2mm}
    \begin{tabular}{ |l| r @{\hspace{2pt}}c@{\hspace{3pt}} l |c| }
        \hline
        RTD Location in Figure~\ref{fig:Thermal_simulation} & \multicolumn{3}{|c|}{Measured Range ($^{\circ}$C)} & Simulation ($^{\circ}$C) \\
        \hline
        1: Near top of outer vessel, opposite cameras & \qquad12.9 & $\pm$ & 0.2 & 12.8 \\
        2: Near middle of outer vessel, opposite cameras & \qquad13.1 & $\pm$ & 0.2 & 12.8 \\
        3: Near bottom of outer vessel, opposite cameras & \qquad11.5 & $\pm$ & 0.2 & 12.8 \\
        4: Outside bottom layer of HDPE insulation & \qquad-14.1 & $\pm$ & 0.2 & -10.9 \\
        5: Stainless-steel base flange & \qquad-10.1 & $\pm$ & 0.2 & -12.0 \\
        \hline
        C$_{3}$F$_{8}$ temperature; min & \multicolumn{3}{|c|}{\,\,-} & 12.8 \\
        C$_{3}$F$_{8}$ temperature; mean & \multicolumn{3}{|c|}{\,\,-} & 12.8 \\
        C$_{3}$F$_{8}$ temperature; max & \multicolumn{3}{|c|}{\,\,-} & 12.9 \\
        \hline
    \end{tabular}
    \label{tab:temperature_comparisons}
\end{table}

\begin{figure}[htbp]
    \centering
    \includegraphics[totalheight=0.40\textheight]{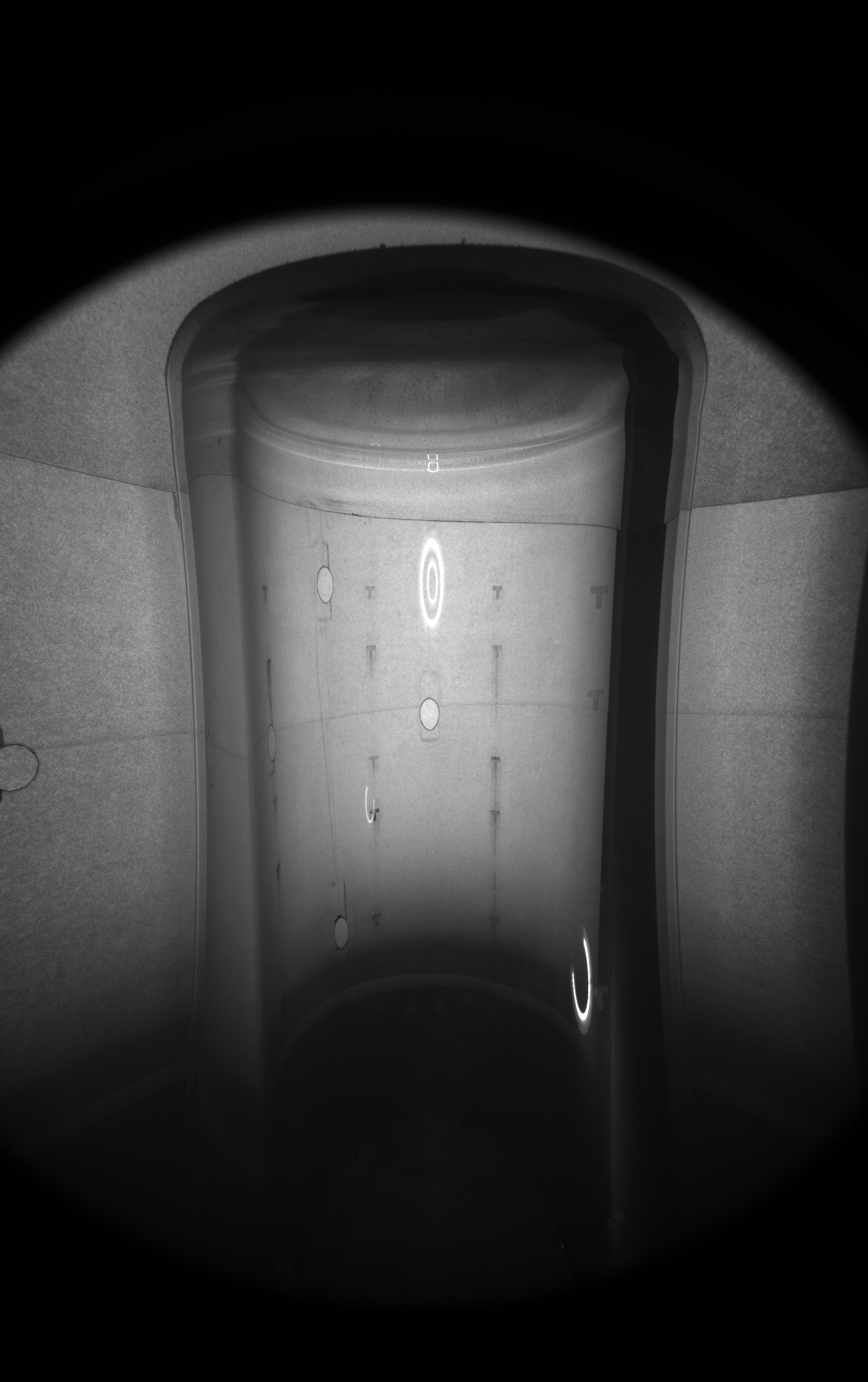}
    \includegraphics[totalheight=0.40\textheight]{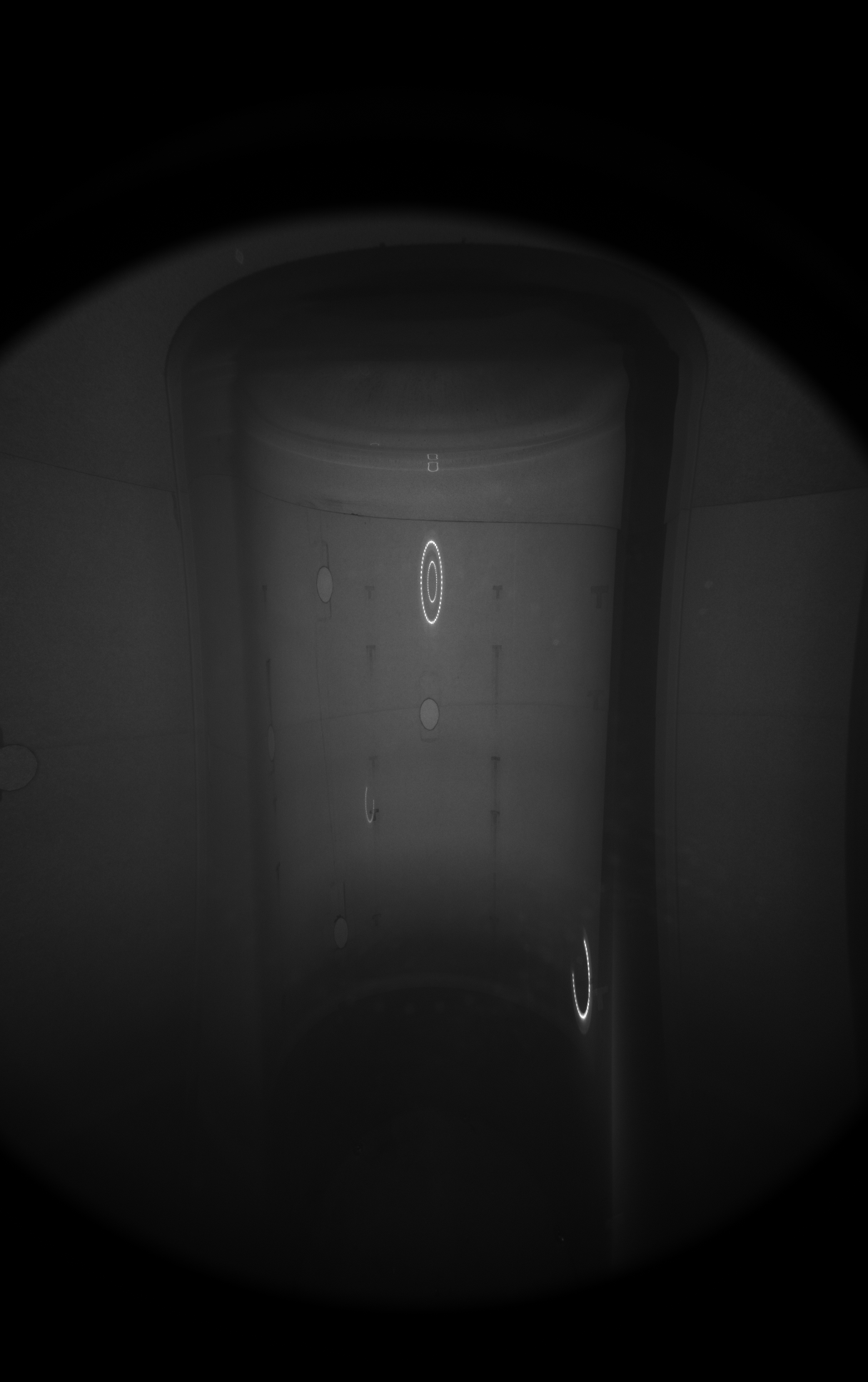}
    \caption{Visual comparison of the temperature impact on optical clarity. Both images were taken within one month of each other with the same LED brightness. (Left) Oil temperature is \SI{18}{\degreeCelsius}. (Right) Oil temperature is \SI{12}{\degreeCelsius}.}
    \label{fig:transparency}
\end{figure}

Temperature stability, while necessary to achieve physics goals, also impacts the optical clarity of the mineral oil through which images of the active volume are captured. Figure~\ref{fig:transparency} gives a visual comparison of two camera images taken with an oil temperature of \SI{18}{\degreeCelsius} (\SI{5}{\degreeCelsius} higher than the nominal operating temperature, left) and \SI{12}{\degreeCelsius} (\SI{1}{\degreeCelsius} lower than the nominal operating temperature, right). The images were taken within one month of each other and with the same LED brightness settings.

\subsection{Camera system and optical trigger}
\label{sec:cameras_and_illumination}
PICO-40L has an optical system with four monochrome Basler ace acA1920-155um cameras paired with Kowa LMVZ4411 varifocal lenses observing the inner detector under a nominal angle of \SI{14}{\degree} above horizontal and \SI{14}{\degree} below horizontal. Two rows of cameras allow stereo reconstruction with an opening angle of \SI{60}{\degree}. The cameras are each enclosed in a water-tight enclosure with an LED ring and heat exchanger. Thermal sensors are also present on the LED rings. Each ring holds 48 red LEDs (model: XPEPHR-L1-0000-00801 from Cree) and is equipped with a light-diffusing foil and a seal to prevent stray illumination from the LEDs from directly entering the camera aperture. The cameras are mounted on an adjustable mount that enables the camera to move along the view port axis. A linear slide allows the camera to be retracted from the measurement position when adjusting the lens settings. The mount also allows the adjustment of the angle to the PV view port axis with two degrees of freedom. The LEDs receive staggered pulses using six Mightex SLC-AA04-US LED driver boards triggered by the exposure gate from the respective cameras, allowing a view of the detector without reflections from neighbouring camera lights. There is a malfunctioning component in the LED-Mightex system that prevents the lower right camera (as viewed from behind the cameras looking toward the centre of the detector) from operating in this mode, and therefore the LEDs around this camera are always on throughout operation. The camera housings are flushed with dry air to prevent any condensation on the optical surfaces. A humidity sensor is installed in each of the two conduits to the camera housings. The camera equipment is operated at a temperature slightly below the water tank system to avoid introducing thermal instability to the inner detector through the view ports.

The back of the PV and the retroreflector cones are covered by embedded glass bead retroreflecting foil (3M Scotchlite Engineer Grade Reflective Sheeting, 3290). The foil in these regions reflects the diffused LED light back towards the cameras, giving a more uniform illumination of the active volume.

To identify an event and trigger the recompression of the detector, software threads examine incoming images from each camera to monitor for the appearance of spikes in the information content of the image. The information is quantified by the image entropy of the absolute difference of sequential frames
\begin{align}
\label{eqn:entropy}
    \mathrm{S}_\mathrm{I} = -\sum_i{P_i \log_2 P_i},
\end{align}
where $P_i$ is the fraction of pixels populating intensity bin $i$ of the difference-map histograms generated from consecutive frames~\cite{entropy}. The entropy exceeding a configurable threshold in any camera is sufficient to trigger an event and recompress the target fluid.

\subsection{Acoustic sensors}
\label{sec:acoustic_sensors}
The inner vessel assembly which houses the target fluid is instrumented with twelve piezoelectric transducers to record the acoustic signal generated during bubble nucleations. The location of the transducers is shown in Figure~\ref{fig:Detector}. The transducers are arranged in three layers of four near the base of the outer vessel, inside HDPE rings cut to house individual transducers at specific locations. They are pressed against the outer vessel wall by a spring compressed to \SI{22}{\newton} in contact with the HDPE ring. The transducers in each row are offset from each other by \SI{90}{\degree}, and each row is offset relative to the others by \SI{30}{\degree}, resulting in uniform transducer coverage in $\phi$. The piezoelectric element itself is a \SI{4.9}{\mm}-diameter cylinder of niobium-doped lead zirconate titanate (PZT), with both ends of each cylinder coated with silver conductive epoxy from MG Chemicals (catalogue number 8331S-15G) to ensure good electrical conductivity. To mitigate a source of backgrounds caused by alpha decays of \potwoten{}, the cylinders are manufactured using pre-screened PbO with low \pbtwoten{} content. Ten of the cylinders have the standard length of \SI{8.7}{\mm}, while two have a much shorter length of \SI{1.0}{\mm}, enhancing the transducer response at higher frequencies at the cost of slightly diminished overall sensitivity.

\begin{figure}[htbp]
    \centering
    \includegraphics[width=\textwidth]{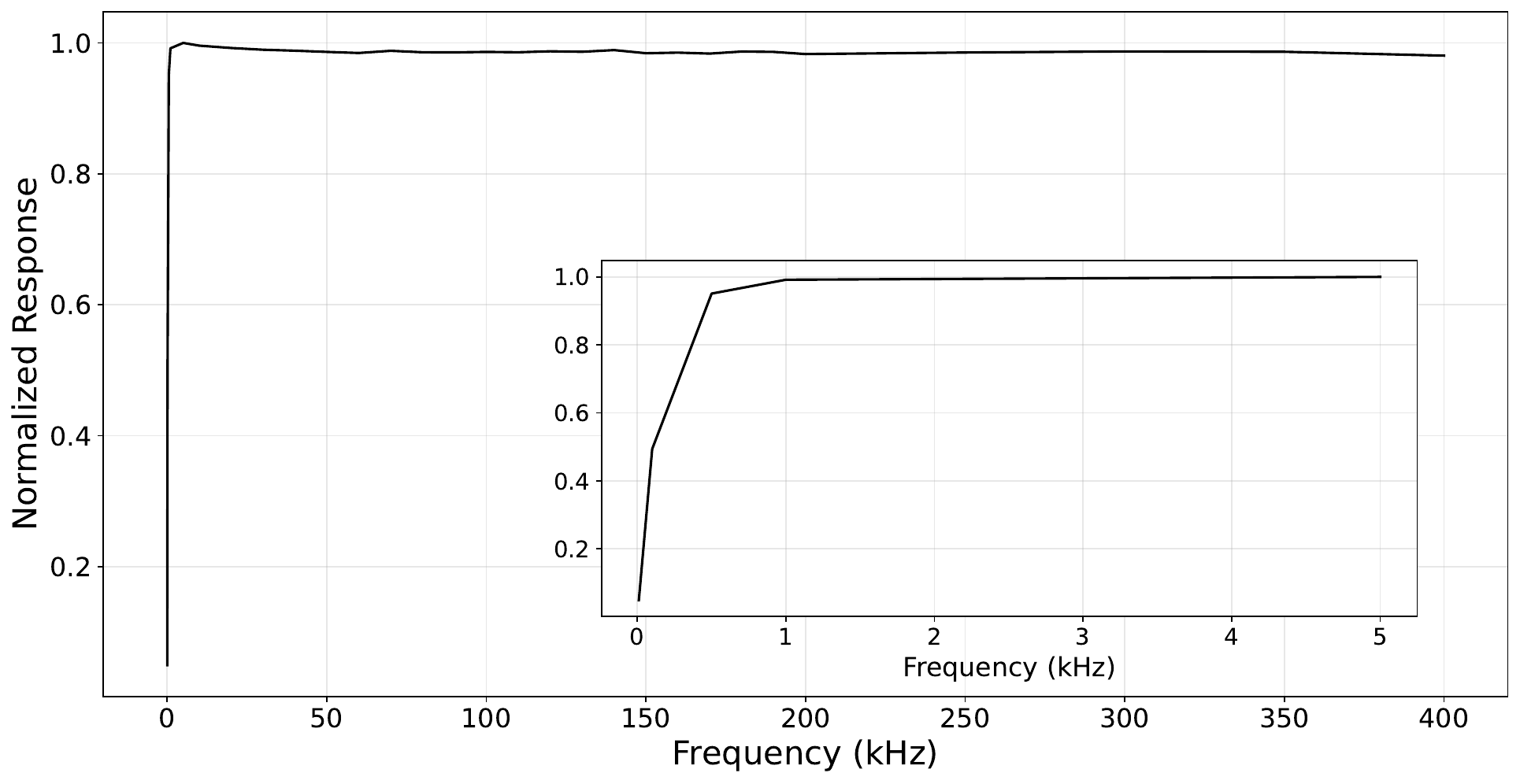}
    \caption{Piezoelectric transducer preamplifier frequency response curve (normalized). The rise at low frequency is shown more closely in the inset plot on the bottom right. The response is flat over the entire frequency range used for the acoustic analysis.}
    \label{fig:Preamplifier_response}
\end{figure}

The electrical signal from the piezoelectric cylinder is fed into a custom preamplifier which is built on a flexible low-radioactivity Kapton printed circuit board (PCB) and located inside the transducer enclosure to minimize electrical noise and interference. The preamplifier has two amplification stages, each with a gain factor of 15.2, with a high-pass filter between them to filter out the DC-offset and low frequency noise below \SI{1}{\kHz}, which is generally caused by other components of the detector and activities in the lab, and is not used in the acoustic analysis. Figure~\ref{fig:Preamplifier_response} shows the frequency response of the preamplifier. Above \SI{1}{\kHz}, the response is flat up to \SI{400}{\kHz}, which encompasses the entire frequency range used in the acoustic analysis and to which the piezoelectric cylinders are sensitive.

\begin{figure}[htbp]
    \centering
    \includegraphics[totalheight=0.3\textheight]{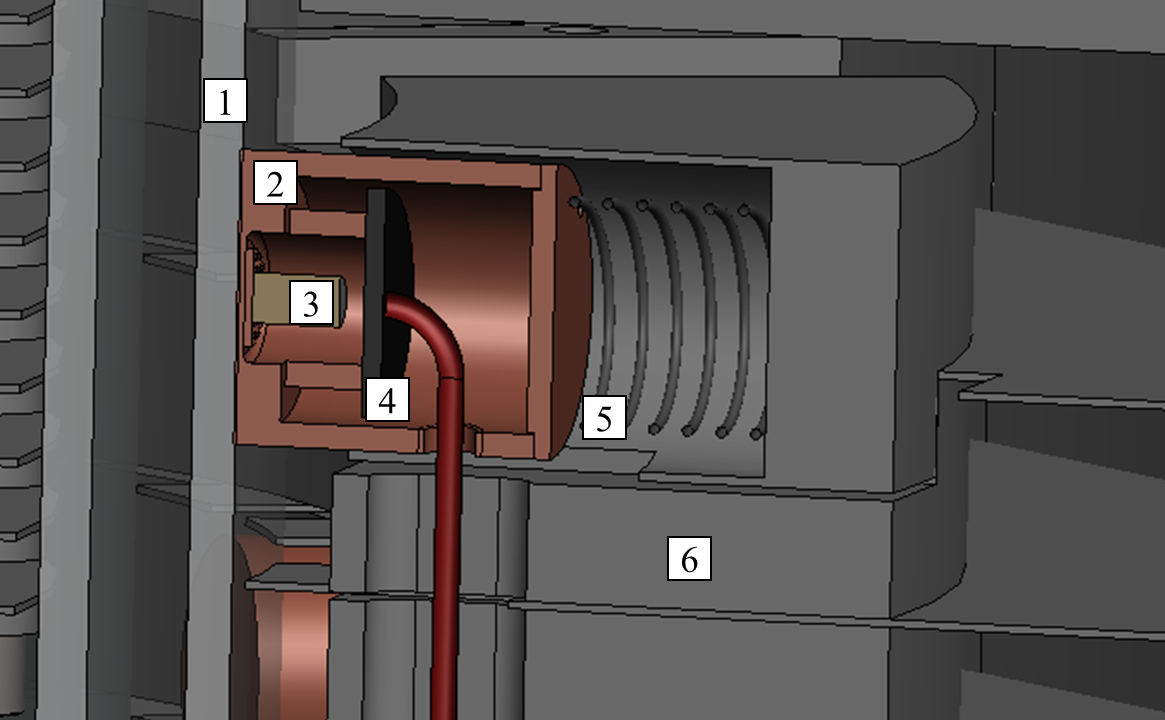}
    \caption{Rendering of a transducer assembly with a standard length (\SI{8.7}{\mm}) piezoelectric cylinder, along with the surrounding infrastructure. The labelled components are (1) the outer quartz vessel wall, (2) copper enclosure, (3) piezoelectric cylinder, (4) preamplifier, (5) \SI{22}{\newton} loading spring, (6) HDPE support ring.}
    \label{fig:PiezoAssembly}
\end{figure}

A labelled rendering of the piezoelectric transducer assembly in situ is shown in Figure~\ref{fig:PiezoAssembly}. The piezoelectric cylinder and preamplifier are housed in an enclosure fabricated from oxygen-free high-conductivity copper, with a diameter of \SI{28.58}{\mm} and a maximum thickness of \SI{17.32}{\mm}. The curvature of the inner face of the enclosure matches the curvature of the outer vessel. Eight of the ten standard-length transducers, and both of the short-length transducers, are backfilled with slow-hardening low-viscosity epoxy from MAS~\cite{MAS} to enhance the overall frequency response of the transducer. The remaining two standard-length transducers have the epoxy only applied to the preamplifier for electrical insulation, with the rest of the enclosed space left empty on installation and ultimately filled with mineral oil as the PV is filled with mineral oil. This latter configuration was found to enhance the response by around a factor of two for frequencies below \SI{50}{\kHz}. All twelve enclosures are covered at the back with a removable copper cap.

\begin{figure}[htbp]
    \centering
    \includegraphics[width=\textwidth]{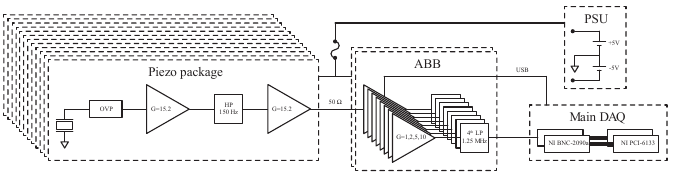}
    \caption{High-level schematic of the processing chain of the piezoelectric transducer signal. The ``Piezo package'' contains the piezoelectric cylinder and preamplifier and is housed in the copper enclosure. The ABB supplies power to the preamplifier and provides additional amplification stages. The DAQ records the amplified signals for offline analysis.}
    \label{fig:Signal_chain}
\end{figure}

A high-level schematic of the signal chain is shown in Figure~\ref{fig:Signal_chain}. The preamplifiers are connected to the Acoustic Breakout Box (ABB) which supplies power to the preamplifiers and provides individually programmable gain factors of 1, 2, 5, or 10 for each channel. After the amplification stage, the signal is passed through a 4th order low-pass Butterworth filter with a cut frequency of \SI{1.25}{\MHz}. The amplified and filtered signal is then digitized using two National Instruments PCI-6133 cards installed in a custom workstation, typically sampling a 14-bit depth at \SI{1}{\mega\hertz} simultaneously on each of sixteen independent channels. The sample clock is shared between devices using a physical NI RTSI cable, similar to a PXI backplane. Acoustic signals are run on BNC cables to two NI BNC-2090A front panels wired in non-referenced single-ended configuration. Each group of eight signals then continues to one of the two digitizer cards along an NI SHC68-68-EPM ribbon cable. Data is acquired in a \SI{32}{MSample} FIFO hardware ring buffer that is timestamped and read out in LabVIEW at the time of the master trigger.

The PV is additionally instrumented with two Dytran 2005V pressure sensors~\cite{dytran}, one in each of the top and bottom instrumentation flanges, sampling the dynamic pressure in the mineral oil at a sampling rate of \SI{1.0}{\MHz}. The signals out of these sensors are routed into the same two NI BNC-2090A panels and NI PCI-6133 cards that receive the amplified signal from the piezoelectric transducers, and are buffered and read out on trigger under the same conditions as those signals. 

\subsection{Controls and data acquisition}
\label{sec:DAQ}

The control and data acquisition (DAQ) system can be split into five main components: the main DAQ, the camera DAQ, the pressure PLC, the temperature PLC, and the slow control system. The main DAQ system is structurally identical to that used in PICO-60~\cite{PICO:2015pux}, consisting of a central state-machine-based LabVIEW operator-control interface, containing modules to interface with the camera system, triggering, alarm conditions, and data acquisition from acoustic digitizers. 

The camera DAQ is built around a modular set of two servers controlling one column of two USB~3.0 cameras each. The timing of the servers is synchronized using a \SI{100}{\Hz} pulse train from a counter output on the main DAQ, connected to direct-coupled general-purpose inputs on each of the cameras. Any camera server observing an above-threshold increase in image entropy as described in Section~\ref{sec:cameras_and_illumination} issues a Transistor-Transistor Logic trigger signal to an NI-9403 digital input/output (DIO) module connected to the main DAQ, which responds by broadcasting a master trigger to the camera DAQ to record images and the pressure PLC to initiate compression.

The pressure PLC consists of a National Instruments cRIO-9074 PLC with analogue and DIO modules controlling the detector pressure state. Coded as a state machine, it switches between the compressed state and expanded state, depending on trigger conditions. It monitors all hydraulic fluid and target fluid pressure transducers and the bellows and IV position sensors at \SI{125}{Hz}. It also controls hydraulic and pneumatic valve states, pressure regulators, and the hydraulic pump. Communication between the PLC and the main DAQ or the slow control is done via MODBUS variables over Transmission Control Protocol/Internet Protocol. Triggering is done via DIO lines for a fast and robust approach. The pressure PLC receives triggers from the main DAQ to compress the detector, and can also initiate a trigger based on pressure monitoring. In the expanded state, a Proportional-Integral-Derivative (PID) loop regulates the hydraulic pressure to achieve the selected target fluid pressure for the selected run conditions. 

The temperature PLC consists of a National Instrument cRIO-9035 PLC and a NI-9149 chassis, and is designed to control and monitor temperatures within the detector. It collects the temperature data from the RTDs distributed throughout the detector every two seconds, and controls 4 chillers (via RS-232) and 3 heaters powered by a DC power supply (TDK-Lambda Z60-14), cooling and heating the detector. The temperature PLC also monitors a number of other experimental parameters: the flow, conductivity, temperature, and leak status of the water tank circulation system; the humidity of the camera enclosures; and chiller fluid flows and pressures.

The slow control system runs on a Linux server, with a second server used for redundancy. It consists of a script polling the PLCs through MODBUS variables for slow data (pressures, temperatures, etc.) every two seconds and storing them in an SQL database. An alarm system monitors critical variables and sends alarm messages via SMS and e-mail to operators. A Python-based graphical user interface allows operators to monitor the status of the detector as well as perform low-level debugging, operation of the detector systems, and plotting of real-time data and past data from the SQL-based database. High-level reconstructed and long-term diagnostic information is automatically processed from the raw data within 1--2 hours to ensure ongoing data quality.

\subsection{Calibration systems}

Calibration of the detector response to backgrounds and radiation that mimics a WIMP scattering signal is achieved using radioactive sources. The calibration system is designed to deliver and house the appropriate gamma and neutron sources safely and position them repeatedly and accurately. PICO-40L utilizes two delivery tubes for calibration: the radioactive source delivery tube (shown in Figure~\ref{fig:Detector}), and the water tank source tube (not shown). The radioactive source delivery tube is a stainless-steel tube (\SI{2.54}{\cm} OD, \SI{1.93}{\cm} ID) that resides inside the PV, connected to a size~2 polyvinyl chloride (PVC) tube (\SI{6.03}{\cm} OD, \SI{4.93}{\cm} ID) via a taper flange with the south flange on top of the PV. The water tank source tube is a \SI{3.81}{\cm} Schedule~80 PVC tube (\SI{4.83}{\cm} OD, \SI{3.81}{\cm} ID) wrapped outside the PV on the north-east side. The access points for both tubes are on top of the water tank through PVC bulkhead fittings, capped while not in use to minimize contamination. These tubes stay in place throughout the operation of the detector. The radioactive source delivery tube is specially designed for a \SI{37}{MBq} $^{60}$Co gamma source, deployed using a \SI{7.62}{m} Gardner Bender EFT-21PN cable snake. The water tank source tube can accommodate many other available sources, and is used to house mainly neutron sources.

\section{PICO-40L installation}

\subsection{PICO-40L cleaning}
\begin{figure}[htbp]
    \centering
    \includegraphics[width=0.9\textwidth]{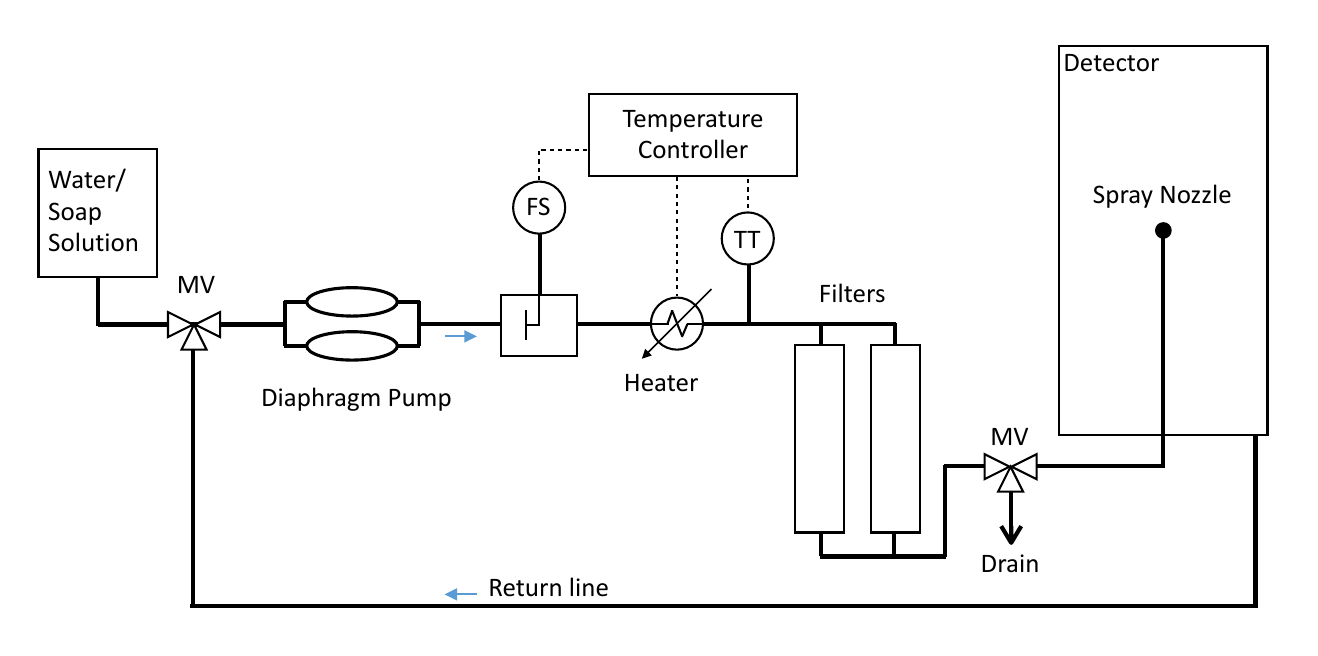}\\
    \caption{Flow path of the cleaning circulation system used for PICO-40L. A \SI{1500}{W} heater enabled the system to reach a water temperature of up to \SI{65}{\degreeCelsius}. The filters used \SI{0.05}{\micro\meter} cartridges. The pump system was connected to clean the inner and outer vessels separately with different spray nozzle setups.}
    \label{fig:cleaning_cart}
\end{figure}

In PICO-2L and PICO-60, particulate contamination was shown to be a concern for stable and low-background operation of bubble chambers~\cite{PICO:2015pux}. To mitigate particulate contamination, a self-contained cleaning circulation system was built to clean the quartz vessels before installation. The circulation system is shown schematically in Figure~\ref{fig:cleaning_cart}. The system was designed with two high-flow filters with \SI{0.05}{\micro\meter} pore size, a \SI{1500}{W} heater capable of bringing the water in the system up to \SI{65}{\degreeCelsius}, a temperature controller, and a \SI{13.6}{L$\mathbin{\cdot}$min$^{-1}$} double diaphragm pump. Rotating spray nozzles and several large-angle nozzles with adjustable elevation were used to spray cleaning solution (a mixture containing Radiacwash and Alcojet) over the inner and outer surfaces of the quartz vessels, respectively. The vessel surfaces were then rinsed with warm UPW until the liquid particulate count of the rinse water met or exceeded the standard set by the Institute of Environmental Sciences and Technology Contamination Control Program (IEST-STD-CC1246D-25)~\cite{cleanliness_standard}. Custom polypropylene stands, designed to only contact the quartz vessels on surfaces that would not be in contact with the target fluid in the final assembly, were constructed to hold and manipulate the vessels during cleaning. The vessels were then dried with boil-off nitrogen before integration with the bellows.

\begin{figure}[htbp]
     \centering
       \includegraphics[width=0.49\textwidth]{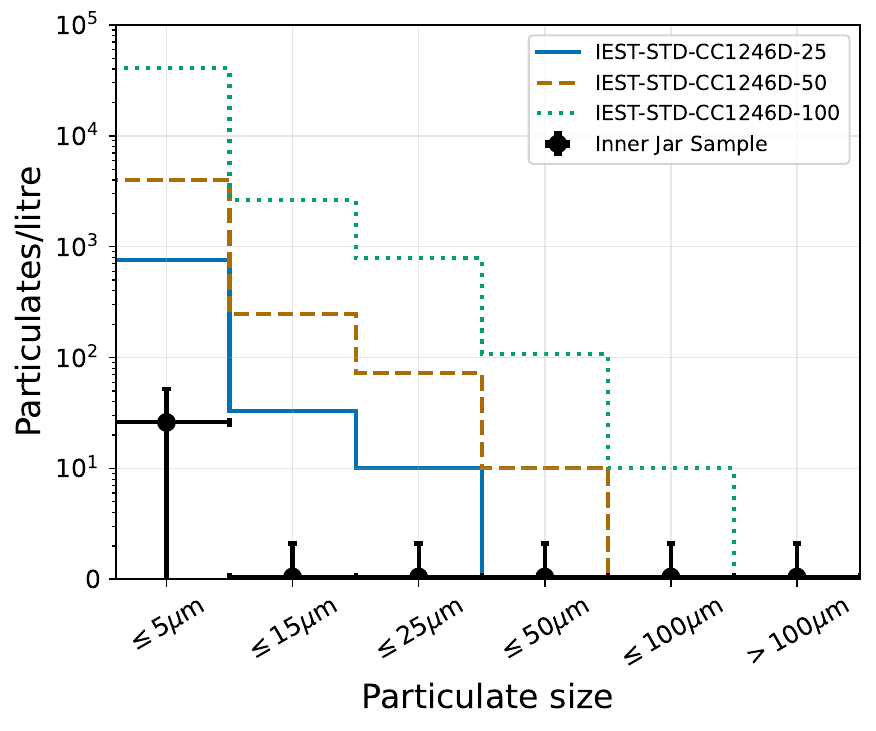}
       \includegraphics[width=0.49\textwidth]{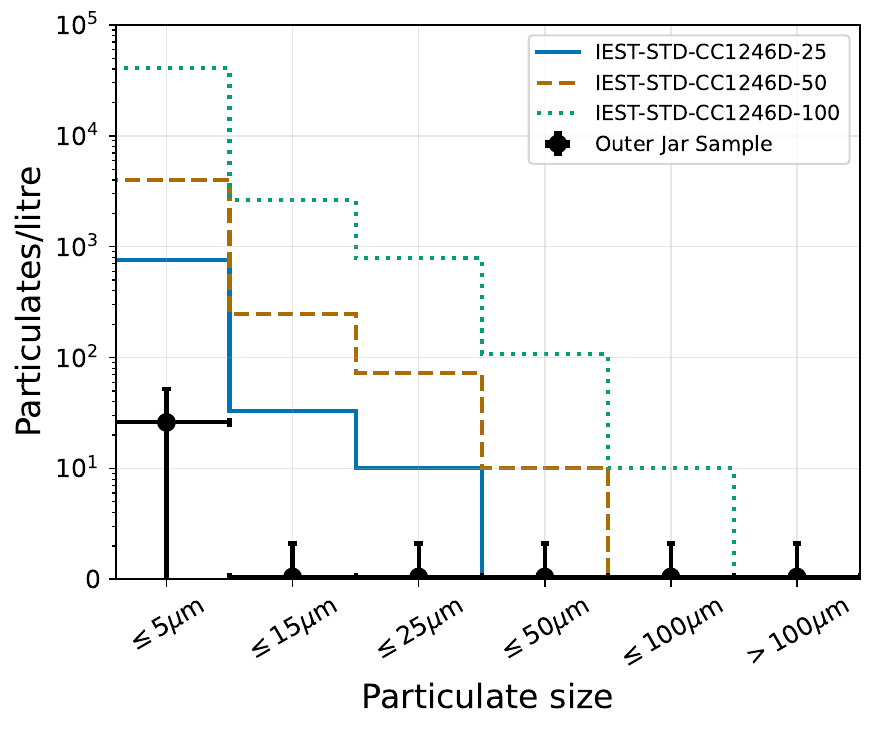}
     \caption{Particulate counts in water samples from the final outer vessel cleaning (left) and final inner vessel cleaning (right). The curves correspond to Institute of Environmental Sciences and Technology product cleanliness level standards~\cite{cleanliness_standard}. The final samples show particulate counts below the strictest standard for all particulate sizes for both vessels, which was a requirement before final assembly.}
     \label{fig:particulates}
 \end{figure}

To evaluate the cleanliness of the vessels, samples of the rinse water after cleaning were obtained from the bottom of the circulation system filter housings. The samples were filtered using \SI{25}{mm}-diameter PTFE membrane filters from Advantec MFS with \SI{0.5}{\micro\meter} pore size. The sizes of the particulates captured on the filters were measured using an optical microscope with a 10x magnification and a \SI{9}{MP} camera. Software was developed to find particulates and automatically size them~\cite{particulate_code}. Particulate counts for samples obtained after the washing and rinsing procedure are shown in Figure~\ref{fig:particulates}. 

\subsection{Assembly and underground installation} 
\label{sec:installation}
The vessels were cleaned and integrated with the bellows assembly in a clean room on surface at SNOLAB. After the vessels were fully attached and sealed, structural supports and a system of seismic support springs were attached to the inner vessel assembly. The space enclosed by the inner vessel assembly was then kept under vacuum to limit radon exposure. The completed assembly was shipped underground in a custom, sealed enclosure. Upon arrival, it was discovered that the inner volume had not held vacuum, and a leak checking campaign was undertaken to identify and remove all leaks. Once repaired, a radon emanation measurement was made of the inner volume and a rate of 20.1 $\pm$ 2.0 atoms of $^{222}$Rn per day was measured, driven by emanation from the nitrile O-ring seal.

The domed top and cylindrical component of the PV (the ``PV body'') was detached from the flat bottom flange of the PV (the ``base flange'') to allow components to be installed separately on each prior to a final integration. The underground assembly of PICO-40L proceeded in six major stages: installation of components to the outside of the PV body, installation of internal components on the inside of the PV body, connection of the inner vessel assembly and external components to the base flange, installation of remaining internal components on the base flange, mounting the PV body to the base flange, and closing the water tank.
 
The camera enclosures were mounted on the outside of the PV body, along with pressure-rated fused silica windows (Corning 7980) into the PV body view ports. The top instrumentation flange, which houses two pressure transducers and a Dytran 2005V fast pressure transducer, was then installed, completing the installation of the external PV body components. The inside wall of the PV was then cleaned and the wall opposite the view ports was covered by the retroreflecting foil. The retroreflector cones, top cooling line, radioactive source deployment tube, and RTDs were then installed inside the PV body, completing the installation of the internal PV body components.

In preparation for the installation of the inner vessel assembly, the three FPAs were installed and connected to the bottom of the base flange, and a custom HDPE component to thermally insulate the plumbing internals was installed on the top of the base flange. The bottom instrumentation flange, which houses two pressure transducers, a Dytran 2005V fast pressure transducer, and the two Gefran position sensors, was installed, along with an electrical feedthrough flange and cold flange. A heating/cooling assembly, consisting of the inner cooling coil, the inner portion of the heating plate, and layers of insulation, was then installed, on top of which the inner vessel assembly was lowered onto the base flange and secured.

The \cthreefeight{} fill lines were then connected and the HDPE ring which houses the twelve piezoelectric transducers was installed around the base of the outer vessel. Each transducer was seated in the appropriate cavity in the ring, with the front face resting against the vessel surface. After all electrical connections were tested, thermally insulating HDPE blocks were added around the cold region. Nickel-chromium (NiCr) heater wire was embedded into a machined groove on the top surface of the top insulation layer, on top of which the copper plates were then installed in two halves around the inner vessel. The PV body was then lowered onto and bolted to the base flange, completing the mechanical assembly.

\subsection{Oil and target fluid processing}
\label{sec:fluid_fill}
After the detector components were assembled, the PV was helium leak tested before the start of the fluid fill. The inner detector (volume within the inner vessel assembly) and outer detector (volume external to the inner vessel assembly but within the PV) were evacuated, pulling degassed mineral oil through a \SI{10}{\micro\meter} inline filter into the outer detector under continued vacuum pumping. Once the PV was filled with the hydraulic fluid, the inner detector was cooled while the \cthreefeight{} supply cylinder was heated. The target fluid evaporated in the storage tank and was filtered through a Pall gas filter (GLP2FCPVMM4) and piped through heated stainless-steel lines into the inner detector, where it condensed. The filling process took place under an operational pressure of about \SI{965}{kPa}, driven by the pressure of the supply cylinder and the heat output of the \cthreefeight{} storage heating system. Once the fill was complete, the detector was compressed to \textasciitilde\SI{1310}{kPa}, preparing the detector for cooldown and data taking.

\section{Backgrounds and shielding}
Background events in PICO-40L which have the potential to mimic a dark matter signal may be broadly divided into two categories: nuclear recoil (NR) events and electronic recoil (ER) events. As mentioned in Section~\ref{sec:loc_dc}, the probability of ER events to nucleate a bubble deviates from the Seitz model prediction, allowing the ER event rate to be reduced through selection of the operating temperature and pressure. The nominal operating \setpoint{}s to optimize sensitivity to dark matter interactions while keeping ER events sufficiently suppressed are a hot region temperature of \SI{13}{\celsius}, and an expansion pressure of \SI{189.6}{kPa} in the centre of the active volume. This corresponds to a Seitz threshold of \SI{3.3}{\keV}. Background NR events cannot be suppressed in this way, since any reduction in the probability of NR events to nucleate a bubble would entail a corresponding reduction in WIMP sensitivity. The primary sources of background NR events in PICO-40L are neutron scattering events and coherent elastic neutrino-nucleus scattering (CE$\nu$NS) events. Neutron-induced events are mitigated through careful detector design and materials selection, with the requirement that less than 1 neutron-induced single-bubble event per live-year should be expected. At \SI{3.3}{keV}, the CE$\nu$NS rate is expected to be the dominant single-bubble background. A summary of the expected background rates originating from each source is shown in Table~\ref{tab:backgrounds}.

\begin{table}[htbp]
    \centering
    \caption{Estimated single-bubble background rates per live-year at a Seitz threshold of $3.3\,\mathrm{keV}$. Median values, central 68\% intervals (16th--84th percentiles), and 90th percentile upper limits, are given. Backgrounds are combined via Monte Carlo sampling, where the distribution for the cosmogenic neutron background is assumed to be half normal. We note that a total median constructed via Monte Carlo sampling need not be constrained to the sum of the medians of the individual components.}
    \label{tab:backgrounds}
    \vspace{2mm}
    \begin{tabular}{|l|c|c|}
        \hline
        Source & Events per Live-Year & 90\% Upper Limit \\
        \hline
        CE$\nu$NS & $1.40^{+0.38}_{-0.33}$ & $1.89$ \\
        Radiogenic Neutrons & $0.58^{+0.29}_{-0.21}$ & $0.966$ \\
        ER Events & $0.05^{+0.37}_{-0.04}$ & $0.776$ \\
        Cosmogenic Neutrons & --- & $0.1$ \\
        \hline
        \textbf{Total Singles-Bubble Events} & $\mathbf{2.23^{+0.65}_{-0.47}}$ & $\mathbf{3.18}$ \\
        \hline
    \end{tabular}
\end{table}
\subsection{Monte Carlo simulations \label{monte_carlo}}

A Monte Carlo simulation model has been developed using Geant4~v11.3.2~\cite{geant4} (with the \texttt{Shielding} physics list for particle transport) to assess the background contribution from radiogenic and cosmogenic neutrons, as well as electronic recoil events produced by beta and gamma rays. For each simulated neutron event, each energy deposition in the active volume is considered a candidate for a bubble nucleation. The probability that each candidate causes a bubble in simulation, and therefore whether the event as a whole is considered a single-bubble or multi-bubble event, is taken from nuclear recoil efficiency curves for carbon and fluorine fit to previous PICO calibrations~\cite{PICONR}. Electronic recoils in \cthreefeight{} generally do not deposit sufficient energy within a critical radius to nucleate bubbles in PICO-40L. Instead, each delta electron emitted along its track has some probability of nucleating a bubble via ionization. This probability is modelled by an exponential function
\begin{align}
\label{eqn:nuc_prob}
    \mathcal{P}=A e^{-B f(P, T)},  
\end{align}
where $A$ and $B$ are fit parameters obtained by fitting to gamma calibration data~\cite{PICO:2019rsv}. The function $f$ is proportional to the ionization threshold, which has a different temperature and pressure dependence than the typical Seitz threshold, allowing the ER background rates to be reduced while maintaining the same dark matter sensitivity~\cite{PICO:2019rsv}. In either case, bubbles that are less than \SI{0.25}{mm} apart are combined. This value is chosen because the spatial extent of a single camera pixel is typically at least this large at the outer vessel boundary closest to the cameras, and enlarges further across the active volume.

\subsection{Cosmogenic neutrons\label{subsec:cosmogenic}}

Cosmogenic neutrons are produced by secondary cosmic rays, primarily muons, when they interact with the norite rock or the materials surrounding the detector via inelastic collisions. The very low muon flux \SI{2}{\km} underground at SNOLAB, caused by the natural overburden, greatly reduces this background. Cosmogenic neutrons at SNOLAB have a mean energy of 109 MeV, an energy spectrum extending up to a few GeV, and an estimated flux of $(\phi_n)_{\mathrm{rock}}=5.4 \times 10^{-11}\ n/\mathrm{cm}^2/\mathrm{s}$~\cite{Mei:2005gm}.

In addition to providing shielding from cosmogenic neutrons originating from the rock surrounding the experiment site, neutrons can also be produced in the water of the 20-tonne water tank housing the experiment, with a similar energy spectrum to those produced in the rock. The number of neutrons produced within a target material depends on the atomic mass of that material, and the functional form of the dependence is well described by a power law~\cite{Mei:2005gm}. Given this model, the number of cosmogenic neutrons expected to be produced in the PICO-40L water tank is $(\phi_n)_{\mathrm{water}}=2.04 \times 10^{-13} n/\mathrm{g}/\mathrm{s}$. Muon-induced neutrons in the rock and water are estimated by our simulations to contribute less than 0.1 single-bubble events per live-year.

\subsection{Radiogenic neutrons\label{subsec:radiogenic}}

Radiogenic neutrons from materials and fluids in the detector are produced by spontaneous fission processes, as well as ($\alpha$, $n$) reactions in light elements from alpha particles emitted in the decay chains of $^{238}$U, $^{235}$U and $^{232}$Th, trace levels of which are present in the detector components. These neutrons have an energy spectrum up to a few MeV. Contributions from ($\alpha$, $n$) reactions can be reduced by positioning detector components away from the active volume as much as possible, carefully selecting detector materials, controlling the ingress of \rntwotwotwo{} during installation, and by purifying the two major detector fluids, mineral oil and \cthreefeight{}, to have a low concentration of uranium and thorium. Intrinsic shielding reduces the flux of neutrons reaching the active volume: the mineral oil reduces the neutron flux from the stainless-steel PV and components sitting outside of it, such as the cameras, lenses, LED illumination, and electronics, while the thermal insulation reduces the background in the active volume by providing shielding against neutrons produced in components in the inner vessel assembly. Radiogenic neutrons produced in the norite rock are shielded by the water tank surrounding the experiment.

A careful materials selection was performed with measurements using gamma spectroscopy and Inductively Coupled Plasma Mass Spectrometry. The radioassay program at SNOLAB was developed to determine the concentration of $^{238}$U and $^{232}$Th in critical components near the active volume or with a large mass. Each component has some purity and an associated uncertainty, or is reported as an upper limit. The concentration of \rntwotwotwo{} in air averages \SI{130}{Bq$\mathbin{\cdot}$m$^{-3}$} at SNOLAB, but varies depending on location, air flow, and time of year~\cite{SNOLAB_Radon}. After the PV was sealed and filled, further radon ingress was not possible, and operations were delayed by a few weeks after filling to allow time for the short-lived radon and products to decay. The \cthreefeight{} was purified by distillation and filtration and the mineral oil was verified to have a sufficiently low concentration of uranium and thorium (\SI{<3.92}{mBq/kg} and \SI{<0.31}{mBq/kg}, respectively).  

To obtain an estimate of the expected background contribution from radiogenic neutrons, the yields and energy spectra from radiogenic neutrons are calculated using ALPHANSO~\cite{ALPHANSO} and then propagated in the Geant4 geometries. A 10\% systematic uncertainty in the yields and a 5\% uncertainty on the mass of components is assumed. Uncertainties from the nuclear recoil efficiency function, the component mass, the neutron yields, and the uranium and thorium activities are propagated to the final estimates through Monte Carlo sampling of distributions. In cases where the measured activity is estimated to be below zero (which may happen due to e.g. a downward statistical fluctuation of the background events in the counting apparatus~\cite{snolab_str_2014_004}), a half-normal distribution with a 90th percentile aligned with the reported 90\% upper limit is used for sampling.

Table~\ref{tab:radiogenic_backgrounds} summarizes the estimated backgrounds from all radiogenic neutrons sources for one live-year of data. The main contributions to this class of background are expected to come from the \SI{880}{L} of mineral oil, and the copper heating plate. In total, all detector components are expected to contribute an upper limit of $< 0.966$ single-bubble events from radiogenic neutrons per live-year at a Seitz threshold of \SI{3.3}{\keV}, at a 90\% confidence level.

\begin{table}[htbp]
    \centering
    \caption{Estimated radiogenic neutron single-bubble background events per live-year at a threshold of $3.3\,\mathrm{keV}$. Median values, central 68\% intervals (16th--84th percentiles), and 90th percentile upper limits, are given. Backgrounds are combined via Monte Carlo sampling. We note that a total median constructed via Monte Carlo sampling need not be constrained to the sum of the medians of the individual components.}
    \label{tab:radiogenic_backgrounds}
    \vspace{2mm}
    \begin{tabular}{|l|c|c|}
        \hline
        Component & Events per Live-Year & 90\% Upper Limit \\
        \hline
        Mineral Oil$^\ast$ & $0.28^{+0.27}_{-0.17}$ & $0.642$ \\
        Inner Cu Heating Disk & $0.10^{+0.11}_{-0.10}$ & $0.246$ \\
        Outer Cu Heating Plate & $0.09^{+0.10}_{-0.09}$ & $0.210$ \\
        Piezoelectric Housing & $0.02^{+0.02}_{-0.02}$ & $0.0503$ \\
        Pressure Vessel & $0.023^{+0.015}_{-0.011}$ & $0.0434$ \\
        Lower Retroreflector & $\left(9.7^{+1.2}_{-1.1}\right)\times10^{-3}$ & $0.0113$ \\
        Thermal Insulation$^\ast$ & $\left(4.0^{+3.8}_{-2.5}\right)\times10^{-3}$ & $9.18\times10^{-3}$ \\
        Cameras & $\left(3.9^{+0.5}_{-0.4}\right)\times10^{-3}$ & $4.49\times10^{-3}$ \\
        Inner RTDs & $\left(3.6^{+1.0}_{-1.0}\right)\times10^{-3}$ & $5.00\times10^{-3}$ \\
        Upper Retroreflector & $\left(1.80^{+0.21}_{-0.20}\right)\times10^{-3}$ & $2.08\times10^{-3}$ \\
        Outer Quartz Jar & $\left(9.6^{+4.6}_{-4.5}\right)\times10^{-4}$ & $1.56\times10^{-3}$ \\
        Piezoelectric PCB & $\left(7.4^{+10}_{-2.3}\right)\times10^{-4}$ & $2.10\times10^{-3}$ \\
        Outer RTDs & $\left(2.9^{+0.8}_{-0.7}\right)\times10^{-4}$ & $3.96\times10^{-4}$ \\
        LED & $\left(5.4^{+0.7}_{-0.7}\right)\times10^{-5}$ & $6.34\times10^{-5}$ \\
        Piezoelectric Ceramic & $\left(4.2^{+1.3}_{-1.1}\right)\times10^{-5}$ & $5.96\times10^{-5}$ \\
        Inner Quartz Jar & $\left(3.5^{+1.7}_{-1.6}\right)\times10^{-5}$ & $5.69\times10^{-5}$ \\
        Camera PCB & $\left(9.2^{+18}_{-3.5}\right)\times10^{-6}$ & $3.34\times10^{-5}$ \\
        Lens & $\left(1.4^{+0.6}_{-0.6}\right)\times10^{-8}$ & $2.23\times10^{-8}$ \\
        \hline
        \textbf{Total Radiogenic Neutrons} & $\mathbf{0.58^{+0.29}_{-0.21}}$ & $\mathbf{0.966}$ \\
        \hline
    \end{tabular}
    \par\smallskip
    \begin{minipage}{\linewidth}
        \footnotesize\raggedright $^\ast$Upper-limit assays dominate: the median upper-limit-source sum divided by the median component total is at least $0.75$. Medians and intervals depend on the assumed assay activity distributions.
    \end{minipage}
\end{table}

\subsection{CE\texorpdfstring{$\nu$}{v}NS}
Neutrinos with the potential to generate background CE$\nu$NS events in PICO-40L are primarily $^{8}$B solar neutrinos, with a very small contribution from helium-proton fusion (hep) neutrinos. The total fluxes for each of these were obtained from a recent global fit~\cite{GonzalezGarcia2024SolarFluxes}, with $\Phi_{^{8}\mathrm{B}} = (5.20 \pm 0.10)\times10^{6}\,\mathrm{cm^{-2}\,s^{-1}}$ used for $^{8}$B solar neutrinos and $\Phi_{\mathrm{hep}} = (3.0^{+0.9}_{-1.0})\times10^{4}\,\mathrm{cm^{-2}\,s^{-1}}$ used for hep neutrinos. We compute the expected rate using the $^{8}$B spectrum shape from Winter \emph{et al.}~\cite{Winter2006} and the predicted hep spectrum shape from Bahcall~\cite{Bahcall1997}. Because the energy deposition mechanism for neutrino-induced nuclear recoils is identical to neutron-induced recoils, we apply the same nucleation efficiency functions used in the neutron background estimate. Propagating the uncertainties from these efficiency functions gives an expected $1.40^{+0.38}_{-0.33}$ single-bubble events per live-year from CE$\nu$NS.

\subsection{ER events\label{subsec:ER}}

To obtain an estimate of the background contribution from ER events, the gamma flux was measured near PICO-60 with NaI detectors~\cite{Robinson2015,Fustin2012} and propagated through the Geant4 model. The expected event rate was obtained by applying the nucleation probability described in Equation~\ref{eqn:nuc_prob} to the electronic energy deposited in the active volume. Two sources of uncertainty were modelled: a 10\% systematic uncertainty on the measured flux, and the uncertainty of the nucleation model, given by the spread of the fit parameters. Uncertainties were propagated by sampling different toy models, assuming that the fit parameters $\log A$ and $B^{-1}$ were uncorrelated, and Gaussian. The uncertainties are highly asymmetric because of the nature of the ER model fit. At the nominal operating \setpoint{}s in the centre of the active volume, $T = \SI{13}{\celsius}$ and $P = \SI{189.6}{kPa}$, this gives $0.05^{+0.36}_{-0.04}$ single-bubble events per live-year.

In addition to the external gamma flux, gamma and beta rays from radioactivity intrinsic to the detector contribute to the ER background expectation, primarily $^{137}$Cs, $^{60}$Co, $^{40}$K, the $^{235}$U chain, and the upper and lower $^{232}$Th and $^{238}$U chains. To estimate this contribution to the ER event rate, the radiopurity of the detector components were measured using the radioassay program at SNOLAB, and radioisotopes were generated in simulation in each component according to its measured activity. The decay of these isotopes were then propagated through the Geant4 model using the same procedure described in Section~\ref{subsec:radiogenic}. At the nominal operating \setpoint{}s, we expect a negligible number of events per live-year from ER events from detector components.

\begin{table}[htbp]
    \centering
    \caption{Estimated radiogenic ER single-bubble background events per live-year at a temperature of \SI{13}{\degreeCelsius} and \SI{189.6}{kPa}. Median values, central 68\% intervals (16th--84th percentiles), and 90th percentile upper limits, are given. The totals are obtained through Monte Carlo sampling.} 
    \label{tab:er_backgrounds}
    \vspace{2mm}
    \begin{tabular}{|l|c|c|}
        \hline
        Component & Events per Live-Year & 90\% Upper Limit \\
        \hline
        External (SNOLAB Flux) & $0.05^{+0.36}_{-0.04}$ & $0.740$ \\
        Pressure Vessel & $\left(8.4^{+62}_{-7.5}\right)\times10^{-4}$ & $1.32\times10^{-2}$ \\
        Outer Quartz Jar & $\left(5.9^{+44}_{-5.3}\right)\times10^{-4}$ & $9.13\times10^{-3}$ \\
        Mineral Oil & $\left(5.0^{+38}_{-4.5}\right)\times10^{-4}$ & $7.98\times10^{-3}$ \\
        Lower Retroreflector & $\left(1.1^{+8.1}_{-1.0}\right)\times10^{-4}$ & $1.69\times10^{-3}$ \\
        Camera & $\left(5.7^{+42}_{-5.1}\right)\times10^{-5}$ & $8.94\times10^{-4}$ \\
        Inner and Outer Heating Plates & $\left(4.8^{+36}_{-4.3}\right)\times10^{-5}$ & $7.48\times10^{-4}$ \\
        Upper Retroreflector & $\left(3.8^{+28}_{-3.4}\right)\times10^{-5}$ & $5.87\times10^{-4}$ \\
        Inner Quartz Jar & $\left(2.8^{+21}_{-2.5}\right)\times10^{-5}$ & $4.43\times10^{-4}$ \\
        Thermal Insulation & $\left(1.3^{+9.6}_{-1.2}\right)\times10^{-5}$ & $2.02\times10^{-4}$ \\
        Piezoelectric Cu Housing & $\left(6.9^{+53}_{-6.2}\right)\times10^{-6}$ & $1.11\times10^{-4}$ \\
        Piezoelectric PCB & $\left(1.4^{+11}_{-1.3}\right)\times10^{-6}$ & $2.22\times10^{-5}$ \\
        Lens & $\left(4.8^{+36}_{-4.4}\right)\times10^{-7}$ & $7.64\times10^{-6}$ \\
        Piezoelectric Ceramic & $\left(9.6^{+72}_{-8.7}\right)\times10^{-8}$ & $1.49\times10^{-6}$ \\
        \hline
        \textbf{Total ER} & $\mathbf{0.05^{+0.37}_{-0.04}}$ & $\mathbf{0.776}$ \\
        \hline
    \end{tabular}
\end{table}
\section{Detector performance}
\label{sec:detector_performance}

For PICO-40L to perform as designed and achieve the intended science goals, it must minimally be capable of discriminating between signal and background events, and be operated stably over long time scales. This section describes the detector performance from data obtained during two data-taking runs: the first was a calibration run with a $^{252}$Cf neutron source deployed in the outer source tube, which took place from 28 October 2023 to 2 November 2023, in which a total of 1769 successful expansions were recorded, and the second was a commissioning run with no source deployed, which took place from 2 November 2023 to 7 February 2024, in which a total of 24,831 successful expansions were recorded. For both runs, the detector was operated at a hot region temperature \setpoint{} of \SI{13}{\degreeCelsius} and a pressure \setpoint{} of \SI{189}{kPa}, corresponding to a threshold at the centre of the detector of \SI{2.8}{keV}. We note that this configuration is slightly below the region for which the detector was designed to operate stably, shown in Figure~\ref{fig:parameter_space_kpa}. These data were obtained during a commissioning period when the stability of the cold region of the detector was being investigated, and it was found that this combination of \setpoint{}s, while resulting in an slightly elevated rate of boiling in the bellows region, nevertheless allowed for continuous operation.

\subsection{Data selection and quality cuts}
Events must pass several basic quality cuts before they can be included for analysis. The detector must remain in the expanded state for a minimum time interval before triggering, and upon a trigger the measured inner and outer temperatures of the upper superheated thermal zone must be within a specified range of the respective temperature \setpoint{}s. The measured pressure of the target fluid must be within a specified range of the pressure \setpoint{}. The range requirements for each \setpoint{} were chosen based on the performance of the PID loops holding the detector at that \setpoint{}, and the minimum time interval requirement was chosen based on the observed decay of a transient high-amplitude pressure oscillation immediately after reaching the expanded state. Table~\ref{tab:cut_flow} shows the progressive fraction of events that pass all preceding quality cuts for the calibration and commissioning runs.

\begin{table}[htbp]
    \centering
    \caption{Table outlining the basic quality cuts used to select events for later analysis. For each data-taking run, the row value gives the total number of events that pass the selection cut in that row and all rows preceding it. The corresponding percentage of the total events for that run is given in the parentheses. A total of 1769 and 24,831 successful expansions were recorded during the calibration and commissioning runs, respectively.}
    \vspace{2mm}
    \begin{tabular}{| r @{\hspace{3pt}}c@{\hspace{3pt}} l |p{3.8cm} | p{3.9cm}| }
        \hline
        \multicolumn{3}{|r|}{Cut Description} & Calibration\newline(Percentage) & Commissioning\newline(Percentage) \\ 
        \hline
        Inner heating plate $\lvert \mathrm{T} - \mathrm{T_{set}} \rvert$ & $<$ & \SI{0.03}{\degreeCelsius} & 1769 (100.00\%) & 24,819 (99.95\%) \\
        Outer heating plate $\lvert \mathrm{T} - \mathrm{T_{set}} \rvert$ & $<$ & \SI{0.08}{\degreeCelsius} & 1769 (100.00\%) & 24,800 (99.88\%) \\
        \cthreefeight{} pressure $\lvert \mathrm{P} - \mathrm{P_{set}} \rvert$ & $<$ & \SI{6.2}{kPa}        & 1677 (94.80\%) & 24,068 (96.93\%) \\
        Time since $\mathrm{P_{set}}$ reached & $>$ & \SI{12}{s}                                            & 1393 (78.75\%) & 22,528 (90.73\%) \\
        \hline
    \end{tabular}
    \label{tab:cut_flow}
\end{table}

\subsection{Position reconstruction and resolution}
\label{sec:3d_reconstruction}
\begin{figure}[htbp]
    \centering
    \includegraphics[width=0.9\linewidth]{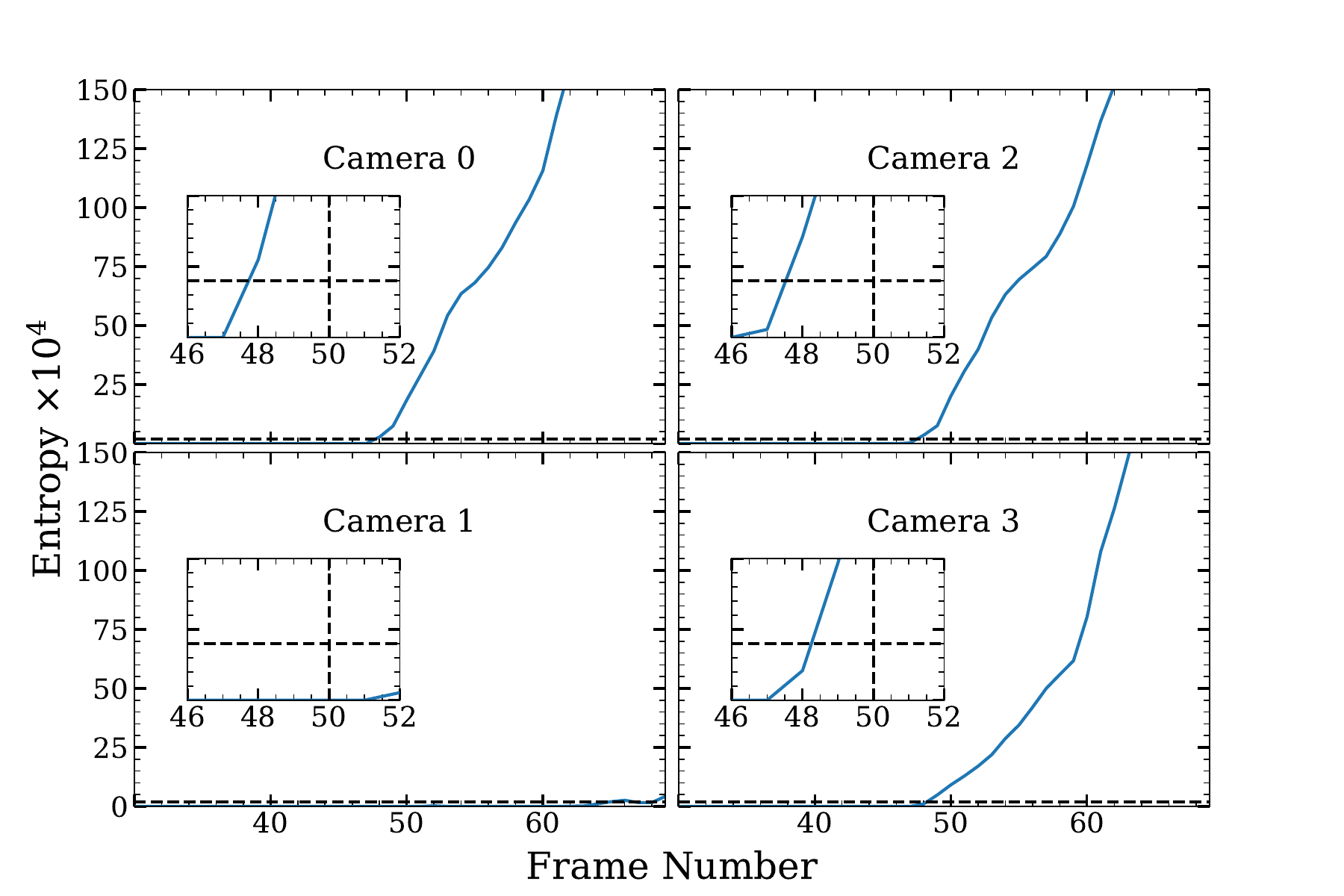}
    \caption{Entropy growth for a typical event in PICO-40L. The horizontal dashed lines correspond to the entropy threshold used to determine the genesis frame. The vertical dashed lines correspond to the trigger frame (which is always frame 50 by construction) Three of the cameras (0, 2, and 3, corresponding to the upper-left, upper-right, and lower-right, as viewed from behind the cameras looking toward the centre of the detector) compute the genesis frame of the bubble to be earlier than the trigger frame (in frames 48, 48, and 49, respectively) while camera 1 is not able to see the bubble until late in the event.}
    \label{fig:entropy}
\end{figure}

The four cameras with which PICO-40L is instrumented allow for the three-dimensional reconstruction of bubble nucleation positions. Upon receiving a trigger, the camera system records the current frame as well as the 20 preceding and 20 succeeding frames. Position reconstruction begins by determining the ``genesis frame'', or the first frame in which a bubble is visible~\cite{Mitra:thesis}. Similar to the technique used for triggering, this process calculates the entropy of the difference between frames according to Equation~\ref{eqn:entropy}. However, instead of calculating the entropy for the difference of consecutive frames, the background image against which the current frame is compared is an average of the first frame of each event in a run. This procedure reduces the noise of the comparison image by exploiting the fact that pre-trigger frames between events are very consistent throughout a run. The frame on which the entropy passes a set threshold is considered the genesis frame for that camera, as shown in Figure~\ref{fig:entropy}.

\begin{figure}[htbp]
    \centering
    \includegraphics[width=0.7\linewidth]{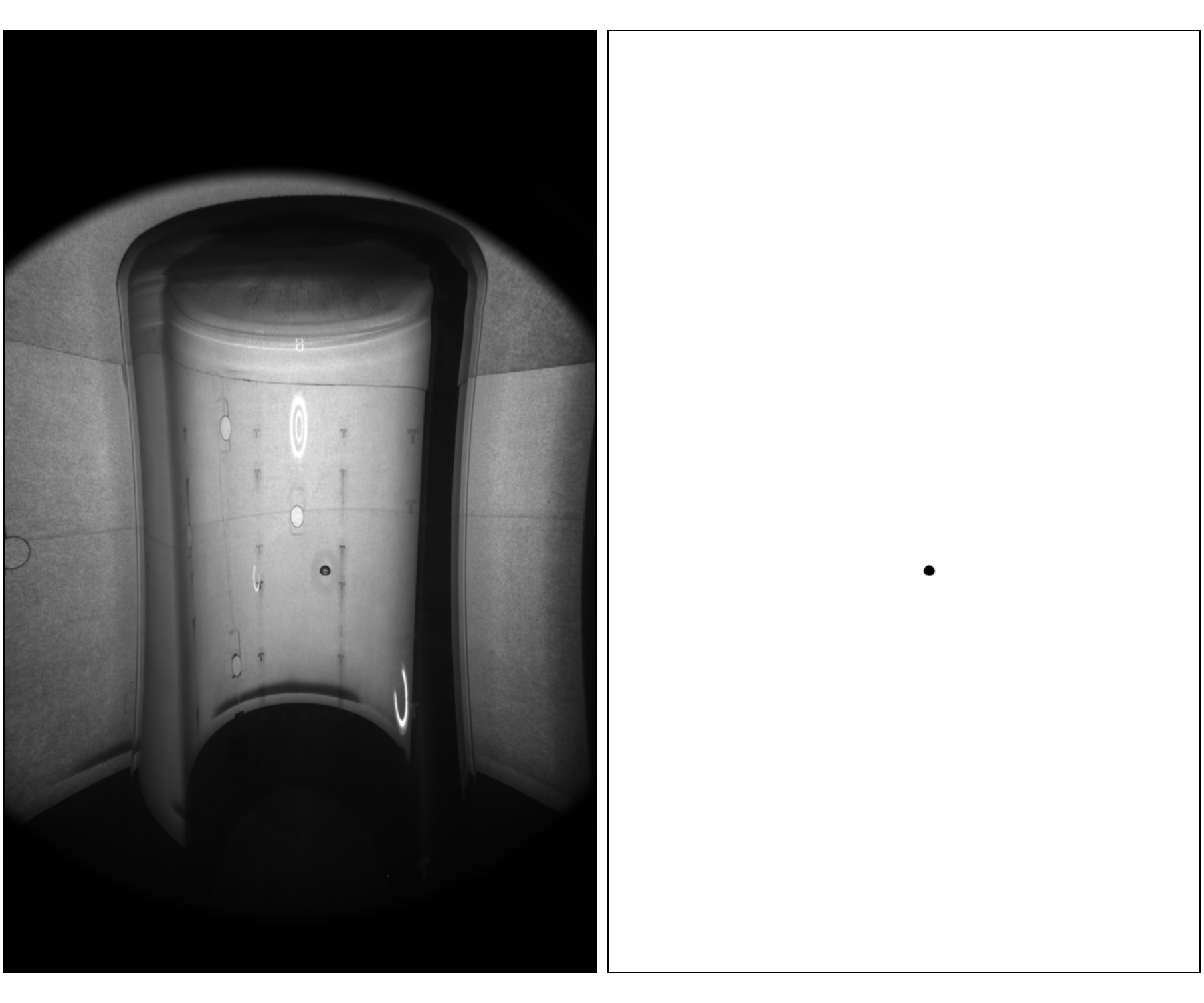}
    \caption{Raw (left) and thresholded difference (right) images from the upper-left camera (camera 0) for a single-bubble event, five frames after the trigger frame, in PICO-40L.}
    \label{fig:im_and_diff}
\end{figure}

Once the genesis frame has been fixed, the location of the bubble(s) in pixel space must be determined. This is accomplished by subtracting the first frame recorded in the event (which shows the detector without bubbles) from the genesis frame, leaving only the bubble in the image. Gaussian smoothing and thresholding are applied to the difference image, which lessens the effect of noise and removes the ``halos'' (faint shadow area surrounding the bubble, which can be seen on the left of Figure~\ref{fig:im_and_diff}) observed during bubble growth. The thresholding process can occasionally leave unfilled holes in the bubble, which are removed using a method known as closing, which is the application of dilation followed by erosion, two common image manipulation techniques~\cite{gonzalez2018digital}. The resulting image after these processing steps is shown on the right of Figure~\ref{fig:im_and_diff}. Finally, the position in pixel coordinates of each bubble is taken as the centroid of the corresponding cluster. This process is repeated for each consecutive frame for all four cameras in order to track bubbles in subsequent frames whenever possible, which has been shown in previous PICO detectors to improve fiducialisation~\cite{PICO:2019vsc}.

After bubble locations in pixel space have been determined, the locations from several cameras are used to reconstruct the 3D position of the bubble(s) within the active volume. The algorithm makes use of a lookup table (LUT) which contains a path through the active volume for each pixel in the image that is capable of imaging the active volume. These paths are generated through the use of a MATLAB-based ray tracing program. This script contains the dimensions and material properties for the detector components relevant to the tracing of light through the active volume. The light rays are traced in reverse, starting at a pixel position on the charge-coupled device within the camera. The ray initially travels away from the optical sensor of the camera, with the direction modified accounting for refraction at the locations of the various surfaces it encounters (view port window, quartz vessel, etc.). For rays that enter the active volume, the entry and exit points are saved and recorded in the LUT for use in the position reconstruction algorithms.

\begin{figure}[htbp]
    \centering
    \includegraphics[width=0.9\linewidth]{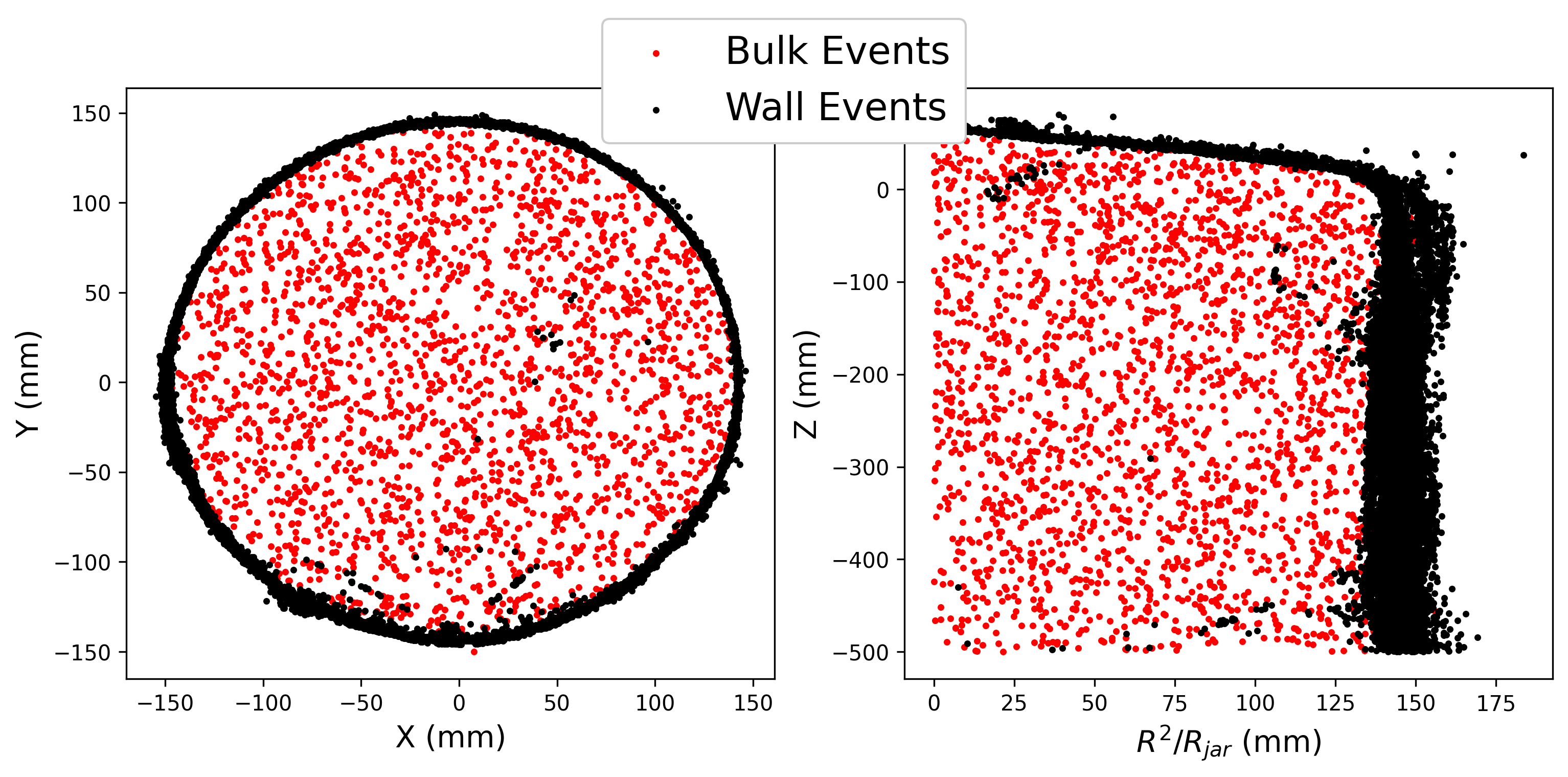}
    \caption[Single-bubble position reconstruction]{Reconstructed position of events that nucleate in the bulk fluid (red) and on the wall (black). To remove events that nucleate on the dome region of the outer vessel, events in the X-Y plot (left) must additionally have a reconstructed $Z$ position within the cylindrical region of the outer vessel. The distinction between bulk and wall events is made using the pressure rise as recorded by the Dytran pressure sensors.}
    \label{fig:alpha_3d}
\end{figure}

Four columns of fiducial marks have been applied to the outer surface of the outer vessel. In addition to constraining the outer vessel position in the ray tracing script, these marks are used to track and correct for apparent shifts in inner vessel assembly position throughout the dataset. Early commissioning data show that the apparent position of inner vessel assembly relative to the cameras is very stable across runs. Figure~\ref{fig:alpha_3d} shows the distribution of reconstructed bubble position during the commissioning run. The distinction between bulk and wall events in the plot is made using an alternative discrimination method using the Dytran pressure sensors, which is described below in Section~\ref{sec:dytran_discrimination}.

\begin{figure}[htbp]
    \centering
    \includegraphics[width=0.9\linewidth]{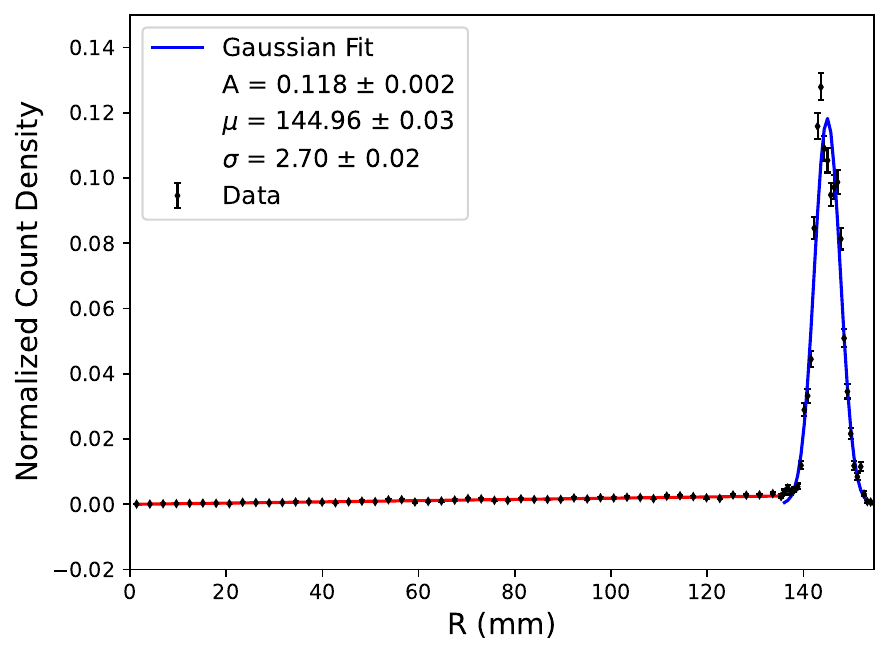}
    \caption[Position resolutions]{Distribution of events in $R$ for the commissioning run. The plot is normalized such that the total integral is unity, and the relative heights of the bulk and wall regions are preserved, compensating for the different bin widths in the different regions. The standard deviation in the Gaussian fit over the wall region is used to estimate the position reconstruction resolution.}
    \label{fig:resolution}
\end{figure}

To obtain an estimate for the position reconstruction resolution, we make use of events where the true $R$ position of nucleation is known, i.e. events that nucleate on the inner surface of the outer vessel housing the target fluid. Since events that occur on the wall of the vessel may be independently obtained using the Dytran pressure sensors, the spread in the reconstructed $R$ position may be used to evaluate the uncertainty in position reconstruction. Figure~\ref{fig:resolution} shows the distribution of events in $R$ for all events obtained during the commissioning run. The region $\pm\SI{10}{\mm}$ around the outer vessel boundary of $\SI{145}{\mm}$ is fit with a Gaussian distribution, and the standard deviation of $\sigma=\SI{2.70\pm0.02}{mm}$ obtained from the fit is taken as an estimate of the reconstruction resolution. Note that the boundary of the outer vessel is where the optical effects of geometrical uncertainties and imperfections in the detector is greatest, and therefore represents a conservative estimate of the true resolution in the bulk of the detector.

\subsection{Bubble time reconstruction}
\begin{figure}[htbp]
    \centering
    \includegraphics[width=\textwidth]{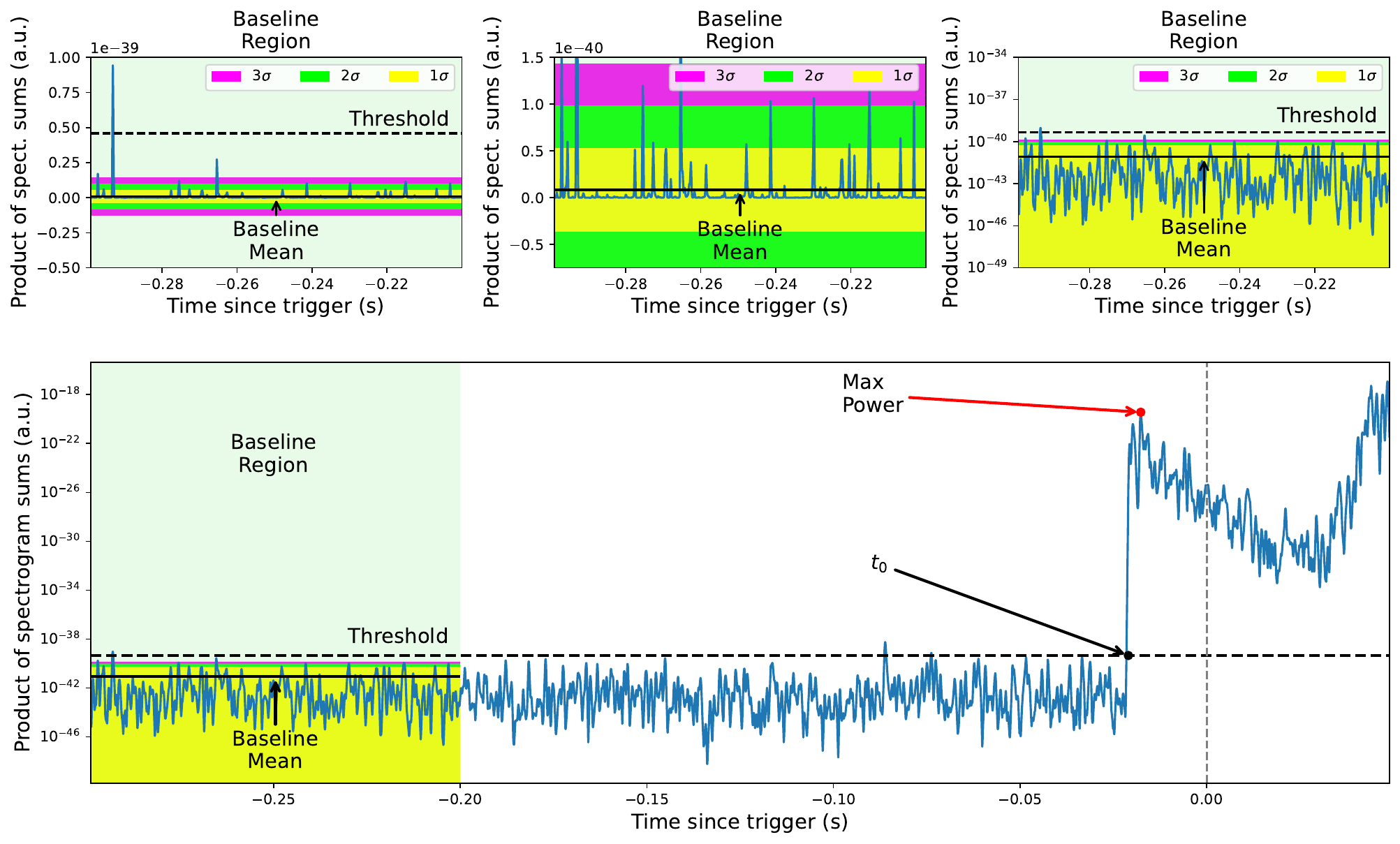}
    \caption{An example signal obtained from applying the bubble time reconstruction algorithm to a piezoelectric transducer signal from a single-bubble event in the bulk of the active volume, including the identification of the acoustic $t_0$. The top-left and top-middle plots show the product of the spectrogram sums in the baseline region before bubble nucleation, including the mean, the bands representing 1, 2, and 3 standard deviations around the mean, and the threshold value of 10 standard deviations above the mean. The top-right plot displays the same information on a log scale, showing the threshold around the upper edge of the noise envelope. The bottom plot shows the full signal, where the sharp rise in signal strength at the time of bubble nucleation is easily identified. $t=0$ is the time of the trigger.}
    \label{fig:t0_finding_example}
\end{figure}

Analysis of bubble growth dynamics for the purpose of event classification, and particularly for discrimination between alpha-decay-induced bubble nucleation and nuclear-recoil-induced bubble nucleation, requires that the start time of the bubble is reliably identified. Figure~\ref{fig:t0_finding_example} shows an example outcome of the algorithm used to determine the bubble start time. A spectrogram of each piezoelectric transducer signal is computed in a window from \SI{300}{\ms} before to \SI{100}{\ms} after the trigger time. Frequency bands where the signal-to-noise ratio is high are selected for each transducer, and the spectrograms within each of these bands are integrated along the frequency axis into a single time series in each band. The product of these bands is then taken to produce a signal which spikes sharply at the time of bubble nucleation. A baseline region is defined from \SI{300}{\ms} to \SI{200}{\ms} before the time of the trigger, well before the beginning of bubble nucleation, and the mean and standard deviation of the signal are calculated for this region. Because the structure of the signal consists of several high-amplitude spikes on a flat baseline, the mean of the baseline region lies just above the flat baseline, and the spikes in amplitude occasionally rise several standard deviations above this mean. This can be seen in the top-left and top-middle plots of Figure~\ref{fig:t0_finding_example}. A threshold is defined at 10 standard deviations above the mean, which typically lies around the upper edge of the noise envelope in log space, as shown in the upper-right plot of Figure~\ref{fig:t0_finding_example}. The time of bubble nucleation (acoustic $t_0$) is identified as one sample after the latest sample in time before the pre-trigger signal maximum that is below the threshold. After this is done for each working transducer, a kernel-density estimate with a Gaussian kernel of width $h =0.03$ is performed to estimate the global $t_0$ maximum, and individual transducer $t_0$ values that lie outside a window around this global $t_0$ are corrected to global $t_0$ before further analysis.

\subsection{Pressure growth discrimination}
\label{sec:dytran_discrimination}
\begin{figure}[htbp]
    \centering
    \includegraphics[width=0.9\linewidth]{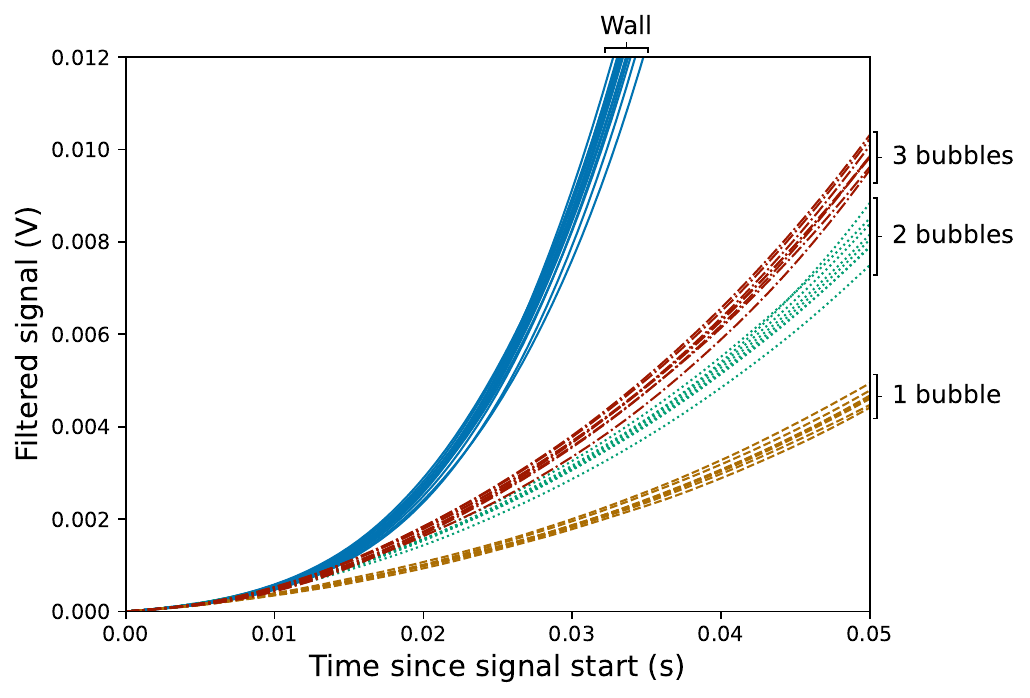}
    \caption{Dytran signal versus time for several events in PICO-40L, illustrating the potential for bubble counting and rejection of wall-nucleated events by pressure rise. For the events shown in the plot, the bubble multiplicity and the location in the bulk or on the wall have been independently confirmed by manual examination of the recorded camera images.}
    \label{fig:dytran_raw}
\end{figure}

\begin{figure}[htpb]
    \centering
    \includegraphics[width=0.49\linewidth]{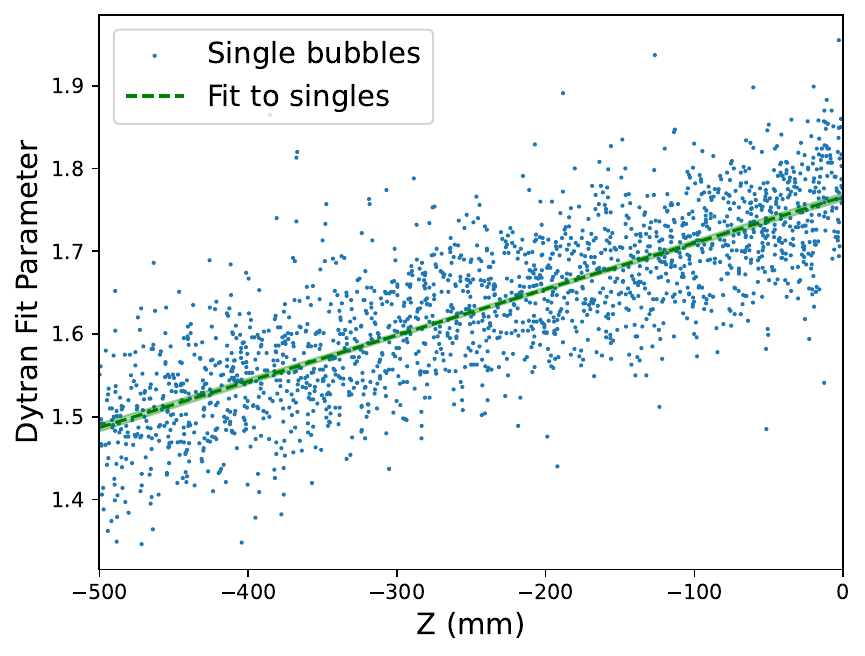}
    \includegraphics[width=0.49\linewidth]{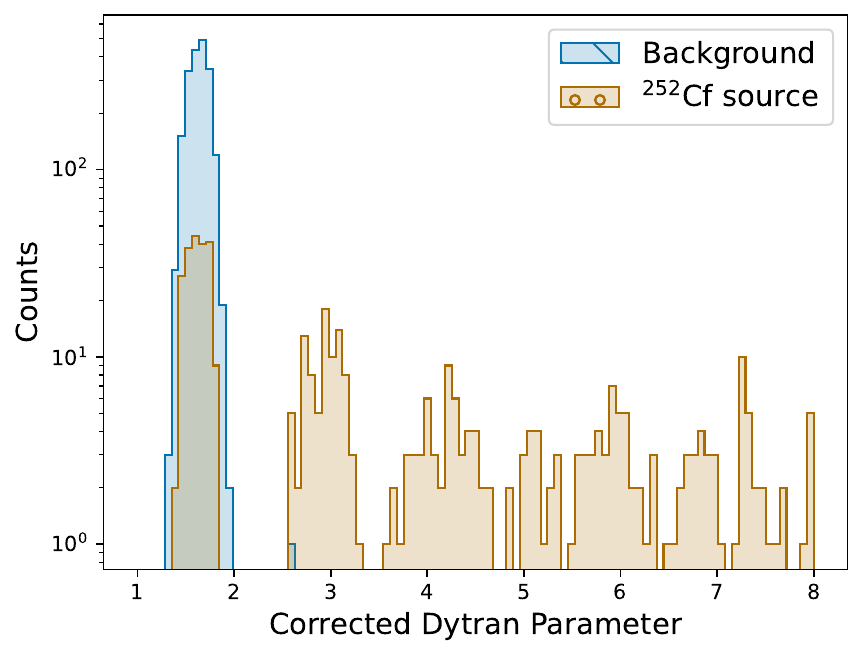}
    \caption{(Left) Fit to single-bubble events in order to correct for hydrostatic pressure. Zero corresponds to the top of the cylindrical region of the outer vessel. (Right) Distribution of the Dytran signal after correction. The single-bubble events are contained in the peak at $x \approx$ 1.7, two-bubble events are contained in the peak at x $\approx$ 3.0, while those at higher values are multi-bubble or wall events.}
    \label{fig:dytran_corr}
\end{figure}

While it is often possible to reject events which contain multiple bubbles or which nucleate close to the wall of the detector using the position reconstruction described in Section~\ref{sec:3d_reconstruction}, certain edge cases may prevent accurate bubble counting, requiring an alternative method. It has been shown in previous PICO detectors~\cite{PICO:2019vsc} that the rate of pressure rise in the mineral oil, read out by the two Dytran 2005V fast pressure transducers, is correlated with the amount of \cthreefeight{} gas produced in the active volume, either from multi-bubble events or nucleation events on the outer vessel wall. This is shown in Figure~\ref{fig:dytran_raw}. To make use of this pressure information, the filtered signal is fitted with a quadratic function over a period of time after the bubble nucleates. The coefficient of the quadratic term is the Dytran fit parameter, and this value is then corrected for hydrostatic pressure using the position determined in Section~\ref{sec:3d_reconstruction}, as shown in the left plot of Figure~\ref{fig:dytran_corr}. The corrected Dytran response can then be used to discriminate single-bubble events from multi-bubble and wall-nucleated events, as shown in the right plot of Figure~\ref{fig:dytran_corr}.

\subsection{Acoustic discrimination}
\label{sec:acoustic_discrimination}
The signal from the twelve piezoelectric transducers are used to produce the acoustic parameter (AP), which allows for discrimination of alpha backgrounds from WIMP-like events~\cite{COUPP:2011}. For each event, a unique signal and noise region of the acoustic trace for each transducer are defined relative to the time at which the bubble nucleated (the acoustic $t_0$). The Fourier transform of each of these regions is scaled by the frequency and subdivided into transducer-specific bins. The scaled frequency content within each bin is summed to produce a value representing the acoustic power contained between the bin edges. The contribution to AP from each transducer $p$ may be defined as
\begin{align}
    \mathrm{AP}_p = \sum_j C^j_p(R^2, Z, \theta) \sum_{f = l}^{u} f \times \mathrm{psd}(f)\,,
\end{align}
where bin $j$ is bounded by frequencies $l$ and $u$, $C^j_p$ is the correction factor at position $(R^2, Z, \theta)$ for frequency bin $j$ and transducer $p$, and psd$(f)$ is the power spectral density of the signal at frequency $f$. The final AP is the sum over the contributions from each transducer.

\begin{figure}[htbp]
    \centering
    \includegraphics[width=0.8\linewidth]{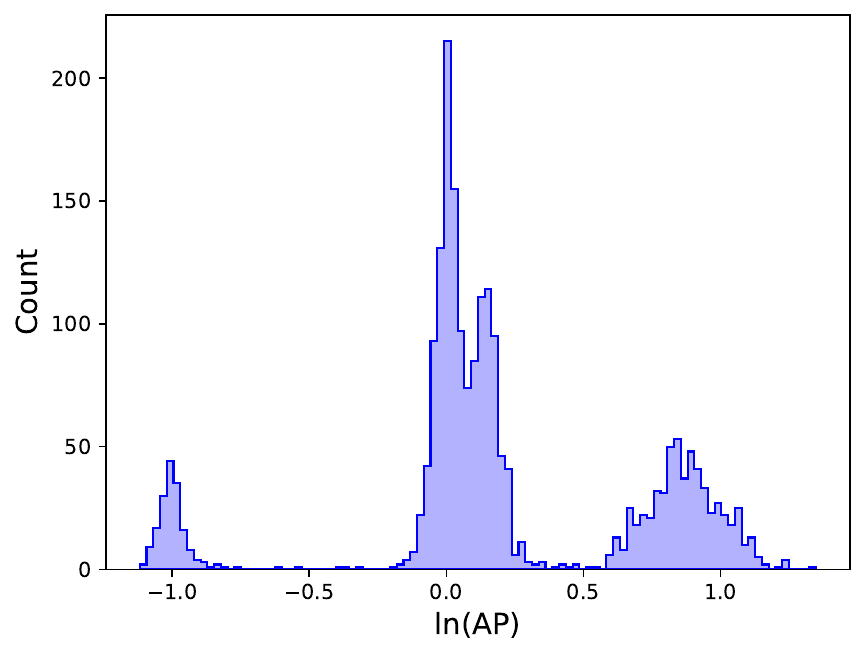}
    \caption{Distribution of the acoustic parameter (AP) for calibration and background data. The leftmost peak (at ln(AP) = -1) is composed of the events from a $^{252}$Cf neutron source, the central pair of peaks (around ln(AP) = 0) is the overlap of the acoustic signal from the decays of $^{222}$Rn and $^{218}$Po, and the highest peak (around ln(AP) = 1) is from the decay of $^{214}$Po.}
    \label{fig:ap-dist}
\end{figure}

A calibration campaign using $^{252}$Cf and $^{60}$Co has been undertaken in order to characterize the response of each transducer and to calculate the position corrections required to produce the AP. The AP distribution for calibration and background data is shown in Figure \ref{fig:ap-dist}. The leftmost peak at around ln(AP) = -1 is composed of events taken during the deployment of a $^{252}$Cf neutron source, and will be used to define the WIMP region of interest. The central pair of peaks around ln(AP) = 0 is the overlap of the acoustic signal from the decays of $^{222}$Rn and $^{218}$Po from the uranium series, while the peak around ln(AP) = 1 is from the decay of $^{214}$Po.

After all selection cuts are applied, a small population of events remains in the WIMP region of interest. A dark matter signal interpretation is rejected by an implied cross-section excluded by several other direct detection experiments, including PICO-60, as well as an observed timing structure of these events relative to decays of \rntwotwotwo{} and daughter isotopes. A current hypothesis for the source of this population is interactions of alpha decays with particulates in suspension in the target fluid causing an acoustic signal that is sufficiently degraded that it enters the WIMP region of interest. Another possibility is the emission of multiple electrons from \pbtwofourteen{} decays in the \rntwotwotwo{} decay chain being sufficiently localized to nucleate a bubble. Further studies are ongoing to isolate the origin of this background.

\section{Summary and outlook}
PICO-40L is the first bubble chamber for dark matter searches that uses the \rightsideup{} design, allowing operation of the detector without a buffer fluid and thereby removing one of the classes of background events that limited earlier generations of PICO detectors. The inverted design requires the introduction of a second quartz vessel to contain the target fluid as well as a new thermal control system to keep the target fluid near the plumbing and instrumentation sub-cooled and suppress bubble nucleation in that region. The optical, acoustic, and fast pressure instrumentation is sufficient to discriminate signal events from events arising from nucleation on the vessel walls, nucleation in the bellows region, and nucleation due to alpha decays, which are the major remaining classes of backgrounds present in the detector. The successful deployment and operation of PICO-40L has demonstrated the potential for large-scale RSU detectors, such as PICO-500 and SBC, to contribute to dark matter searches in the future. We expect that PICO-40L will be fully commissioned in 2026.

\acknowledgments
The PICO collaboration wishes to thank SNOLAB and its staff for support through underground space, logistical, and technical services. SNOLAB operations are supported by the Canada Foundation for Innovation (CFI) and the Province of Ontario Ministry of Research and Innovation, with underground access provided by Vale at the Creighton mine site. We wish to acknowledge the support of the Natural Sciences and Engineering Research Council of Canada (NSERC) and the CFI for funding, and the Arthur B.\ McDonald Canadian Astroparticle Physics Research Institute. We acknowledge that this work is supported by the National Science Foundation (NSF) (Grants No.\ 0919526, No.\ 1506337, No.\ 1242637, and No.\ 1205987), by the U.S.\ Department of Energy (DOE) Office of Science, Office of High Energy Physics (Grants No.\ DESC0017815 and No.\ DE-SC-0012161), by the DOE Office of Science Graduate Student Research (SCGSR) award, by the Department of Atomic Energy (DAE), Government of India, under the Centre for AstroParticle Physics II project (CAPP-II) at the Saha Institute of Nuclear Physics (SINP), and a grant from the Programme Johannes Amos Comenius under the Ministry of Education, Youth and Sports of the Czech Republic [CZ.02.01.01/00/22\_008/0004590]. This work is also supported by SECIHTI Project No. CBF-2025-I-1589 and DGAPA UNAM Grants No. PAPIIT IN105923 and IN102326. This work is partially supported by the Kavli Institute for Cosmological Physics at the University of Chicago through NSF Grants No.\ 1125897 and No.\ 1806722, and an endowment from the Kavli Foundation and its founder Fred Kavli. We also wish to acknowledge the support from Fermi National Accelerator Laboratory under Contract No.\ DEAC02-07CH11359, and from Pacific Northwest National Laboratory, which is operated by Battelle for the U.S.\ Department of Energy under Contract No.\ DE-AC05-76RL01830. We also thank the Digital Research Alliance of Canada (alliancecan.ca) and the Centre for Advanced Computing, ACENET, Calcul Québec, Compute Ontario, and WestGrid for computational support. The work of M.\ Bressler is supported by the Department of Energy Office of Science Graduate Instrumentation Research Award (GIRA).

\bibliographystyle{JHEP}
\bibliography{main_bib}

\end{document}

%% file: authors.tex
\author[a]{E.~Adams,}
\author[b]{B.~Ali,}
\author[c]{R.~Anderson-Dornan,}
\author[d]{D.~J.~Auty,}
\author[e]{E.~Behnke,}
\author[a]{M.~Bai,}
\author[f]{M.~Bressler,}
\author[a]{B.~Broerman,}
\author[a]{G.~Cao,}
\author[g]{W.~C.~Chen,}
\author[a]{U.~Chowdhury,}
\author[a]{K.~Clark,}
\author[h]{J.~I.~Collar,}
\author[a,i]{J.~Corbett,}
\author[d]{C.~B.~Coutu,}
\author[a,i]{D.~Cranshaw,}
\author[j]{M.~Crisler,}
\author[k]{C.~E.~Dahl,}
\author[l]{M.~Das,}
\author[d]{S.~Fallows,}
\author[b]{R.~Filgas,}
\author[c]{A.~Garc\'{i}a-Viltres,}
\author[m]{C.~Gaudreau,}
\author[g,n]{F.~Girard,}
\author[a]{G.~Giroux,}
\author[m]{P.~Grylls,}
\author[a]{C.~Hardy,}
\author[o]{O.~Harris,}
\author[a]{H.~Hawley-Herrera,}
\author[a]{R.~Hill,}
\author[n]{T.~Hillier,}
\author[d]{C.~B.~Krauss,}
\author[g]{M.~Laurin,}
\author[m,n]{I.~Lawson,}
\author[n]{A.~Leblanc,}
\author[p]{H.~Leng,}
\author[e]{I.~Levine,}
\author[d]{Q.~Malin,}
\author[m]{A.~Mathewson,}
\author[p]{D.~Maurya,}
\author[a]{S.~Meister,}
\author[g]{V.~Monette,}
\author[a]{C.~Moore,}
\author[f]{R.~Neilson,}
\author[a]{A.~J.~Noble,}
\author[g]{H.~Nozard,}
\author[h]{A.~Ortega,}
\author[d]{S.~Pal,}
\author[d]{M.-C.~Piro,}
\author[g]{A.~Plante,}
\author[p]{S.~Priya,}
\author[d]{M.~Rangen,}
\author[d]{C.~Rethmeier,}
\author[a]{M.~Robert,}
\author[g]{A.~E.~Robinson,}
\author[g]{J.~Savoie,}
\author[m]{S.~J.~Sekula,}
\author[g]{N.~Starinski,}
\author[b]{I.~\v{S}tekl,}
\author[a]{T.~Sullivan,}
\author[h]{M.~Tripathi,}
\author[c]{E.~V\'azquez-J\'auregui,}
\author[n]{E.~Weima,}
\author[n]{U.~Wichoski,}
\author[d]{W.~Woodley,}

\affiliation[a]{Department of Physics, Queen's University, Kingston, K7L 3N6, Canada}
\affiliation[b]{Institute of Experimental and Applied Physics, Czech Technical University in Prague, Prague, Cz-12800, Czech Republic}
\affiliation[c]{Instituto de F\'isica, Universidad Nacional Aut\'onoma de M\'exico, M\'exico D.\:F. 01000, M\'exico}
\affiliation[d]{Department of Physics, University of Alberta, Edmonton, T6G 2E1, Canada}
\affiliation[e]{Department of Physics, Indiana University South Bend, South Bend, Indiana 46634, USA}
\affiliation[f]{Department of Physics, Drexel University, Philadelphia, Pennsylvania 19104, USA}
\affiliation[g]{D\'epartement de Physique, Universit\'e de Montr\'eal, Montr\'eal, H3C 3J7, Canada}
\affiliation[h]{Enrico Fermi Institute, KICP and Department of Physics, University of Chicago, Chicago, Illinois 60637, USA}
\affiliation[i]{Arthur B. McDonald Canadian Astroparticle Physics Research Institute, Queen's University, Kingston, K7L 3N6, Canada}
\affiliation[j]{Fermi National Accelerator Laboratory, Batavia, Illinois 60510, USA}
\affiliation[k]{Department of Physics and Astronomy, Northwestern University, Evanston, Illinois 60208, USA}
\affiliation[l]{Nuclear Atomic Particle Physics Division, Saha Institute of Nuclear Physics, HBNI, Kolkata, India}
\affiliation[m]{SNOLAB, Lively, Ontario, P3Y 1N2, Canada}
\affiliation[n]{School of Natural Sciences, Laurentian University, Sudbury, P3E 2C6, Canada}
\affiliation[o]{Northeastern Illinois University, Chicago, Illinois 60625, USA}
\affiliation[p]{Materials Research Institute, Penn State, University Park, Pennsylvania 16802, USA}

\collaboration{PICO Collaboration}
\emailAdd{analysis@picoexperiment.com}
\keywords{Dark Matter detectors, Detector design and construction technologies and materials, Large detector systems for particle and astroparticle physics}